\documentclass[a4paper,11pt]{article}
\pdfoutput=1 

\usepackage{jheppub} 

\makeatletter
\DeclareRobustCommand*{\bfseries}{%
  \not@math@alphabet\bfseries\mathbf
  \fontseries\bfdefault\selectfont
  \boldmath
}
\makeatother

\usepackage{calc}
\usepackage{rotating}
\usepackage[english]{babel}
\usepackage{graphicx}
\usepackage{subfig}
\usepackage{float}
\usepackage{amsmath}
\usepackage{amssymb}
\usepackage{amsthm}
\usepackage{latexsym}
\usepackage{dcolumn}
\usepackage{hyperref}
\newcommand{\newc}{\newcommand*}

\long\def\begincomment#1\endcomment{%
        \begingroup\sf\baselineskip12pt#1\endgroup}

\newc{\etal}{\textrm{et al.}} 
\newc{\eg}{\textrm{e.g.}} 
\newc{\ie}{\textrm{i.e.}}
\newc{\etc}{\textrm{etc}}
\newc\vs{\textrm{vs.}}
\newc{\cl}{\rm {C.L.}}

\newc{\ev}{\ensuremath{\,\mathrm{eV}}}
\newc{\kev}{\ensuremath{\,\mathrm{keV}}}
\newc{\mev}{\ensuremath{\,\mathrm{MeV}}}
\newc{\gev}{\ensuremath{\,\mathrm{GeV}}}
\newc{\tev}{\ensuremath{\,\mathrm{TeV}}}
\newc{\MeV}{\mev} 
\newc{\TeV}{\tev}
\newc{\invpb}{\ensuremath{/\text{pb}}}
\newc{\invfb}{\ensuremath{\,\text{fb}^{-1}}}
\newc\nb{\ensuremath{\,\mathrm{nb}}} \newc\pb{\ensuremath{\,\mathrm{pb}}} \newc\fb{\ensuremath{\,\mathrm{fb}}}
\newc\pc{\ensuremath{\,\mathrm{pc}}}
\newc\kpc{\ensuremath{\,\mathrm{kpc}}}
\newc\mpc{\ensuremath{\,\mathrm{Mpc}}}
\newc\ps{\ensuremath{\,\mathrm{ps}}} 
\newc\cmeter{\ensuremath{\,\mathrm{cm}}} 
\newc\meter{\ensuremath{\,\mathrm{m}}} 
\newc\kmeter{\ensuremath{\,\mathrm{km}}}
\newc\second{\ensuremath{\,\mathrm{s}}}
\newc\msecond{\ensuremath{\,\mathrm{ms}}}
\newc\nsecond{\ensuremath{\,\mathrm{ns}}}
\newc\psecond{\ensuremath{\,\mathrm{ps}}}

\newc{\chisqmin}{\ensuremath{\chi^2_{\mathrm{min}}}}
\newc{\Delchisq}{\ensuremath{\Delta\chi^2}}
\newc{\chisq}{\ensuremath{\chi^2}}
\newc{\like}{\ensuremath{\mathcal{L}}}

\newc\lsim{\ensuremath{\mathrel{\rlap{\lower4pt\hbox{\hskip1pt$\sim$}}\raise1pt\hbox{$<$}}}}
\newc\gsim{\ensuremath{\mathrel{\rlap{\lower4pt\hbox{\hskip1pt$\sim$}}\raise1pt\hbox{$>$}}}}
\newc{\VEV}[1]{\ensuremath{\langle #1 \rangle}}
\newc{\dl}{\ensuremath{\stackrel{\leftarrow}{D}}}
\newc{\dr}{\ensuremath{\stackrel{\rightarrow}{D}}}

\newc{\scr}[1]{\ensuremath{\mathcal{#1}}}

\newc{\bcenter}{\begin{center}}    \newc{\ecenter}{\end{center}}
\newc{\bfl}{\begin{flushleft}}    \newc{\efl}{\end{flushleft}}
\newc{\bfr}{\begin{flushright}}    \newc{\efr}{\end{flushright}}

\newc{\bi}{\begin{itemize}}
\newc{\ei}{\end{itemize}}
\newc{\bed}{\begin{description}}
\newc{\eed}{\end{description}}
\newc{\ben}{\begin{enumerate}}
\newc{\een}{\end{enumerate}}

\newc{\be}{\begin{equation}}
\newc{\ee}{\end{equation}}
\newc{\bea}{\begin{eqnarray}}
\newc{\eea}{\end{eqnarray}}
\newc{\ra}{\rightarrow}

\newc{\alphas}{\ensuremath{\alpha_s}}
\newc{\alphatwo}{\ensuremath{\alpha_2}}
\newc{\alphaone}{\ensuremath{\alpha_1}}
\newc{\alphai}[1]{\ensuremath{\alpha_{#1}}}
\newc{\alphaem}{\ensuremath{\alpha_{\mathrm{em}}}}
\newc{\alphaeff}{\ensuremath{\alpha_{\mathrm{eff}}}}
\newc{\sineff}{\ensuremath{\sin \theta_{\mathrm{eff}}}}
\newc{\sinsqeff}{\ensuremath{\sin^2 \theta_{\mathrm{eff}}}}
\newc{\dalphahad}{\ensuremath{\Delta \alpha_{\mathrm{had}}}}
\newc{\yt}{\ensuremath{h_t}} \newc{\yb}{\ensuremath{h_b}} \newc{\ytau}{\ensuremath{h_{\tau}}}
\newc\mz{\ensuremath{M_Z}} 
\newc\mw{\ensuremath{m_W}}
\newc\mZ{\mz}        \newc\mW{\mw}
\newc\mhsm{\ensuremath{ m_{H_{\mathrm{SM}}}}}
\newc{\mtop}{\ensuremath{ m_t}}               \newc{\mtpole}{\ensuremath{ M_t}}
\newc{\mbottom}{\ensuremath{ m_b}} 
\newc{\mtau}{\ensuremath{ m_{\tau}}}
\newc{\mt}{\mtpole}
\newc{\mb}{\mbottom} 
\newc{\rgg}{\ensuremath{R_{h}(\gamma\gamma)}}
\newc{\rzz}{\ensuremath{R_{h}(ZZ)}}
\newc{\rtwogg}{\ensuremath{R_{h_2}(\gamma\gamma)}}
\newc{\rtwozz}{\ensuremath{R_{h_2}(ZZ)}}
\newc{\ronegg}{\ensuremath{R_{h_1}(\gamma\gamma)}}
\newc{\ronezz}{\ensuremath{R_{h_1}(ZZ)}}
\newc{\rsiggg}{\ensuremath{R_{h_\textrm{sig}}(\gamma\gamma)}}
\newc{\rsigzz}{\ensuremath{R_{h_\textrm{sig}}(ZZ)}}
\newc{\llbar}{\ensuremath{\ell\bar{\ell}}}
\newc{\tauptaum}{\ensuremath{ \tau^+\tau^-}}
\newc{\qqbar}{\ensuremath{ q\bar{q}}} \newc{\ppbar}{\ensuremath{ p\bar{p}}}
\newc{\bbbar}{\ensuremath{ b\bar{b}}} \newc{\ttbar}{\ensuremath{ t\bar{t}}}
\newc{\ffbar}{\ensuremath{ f\bar{f}}} \newc{\tautaubar}{\ensuremath{ \tau\bar{\tau}}}
\newc{\mchi}{\ensuremath{m_{\chi}}}
\newc{\squark}{\ensuremath{\tilde{q}}}
\newc{\slepton}{\ensuremath{\tilde{l}}}
\newc{\gluino}{\ensuremath{\tilde{g}}} 
\newc{\mgluino}{\ensuremath{{m_{\gluino}}}}
\newc{\tone}{\ensuremath{{\tilde{t}_1}}}

\newc{\sthw}{\ensuremath{ \sin\theta_W}}              \newc{\cthw}{\ensuremath{\cos\theta_W}}
\newc{\tanthw}{\ensuremath{ \tan\theta_W}}              \newc{\cotthw}{\ensuremath{\cot\theta_W}}
\newc{\ssqthw}{\ensuremath{\sin^2 \theta_W}}
\newc{\msbar}{\ensuremath{\overline{MS}}} \newc{\drbar}{\ensuremath{\overline{DR}}}
\newc{\mtmtsmmsbar}{\ensuremath{ m_t(m_t)^{\msbar}_{{\mathrm{SM}}}}}
\newc{\mtmtsmdrbar}{\ensuremath{ m_t(m_t)^{\drbar}_{{\mathrm{SM}}}}}
\newc{\mtmtmssmdrbar}{\ensuremath{ m_t(m_t)^{\drbar}_{{\mathrm{SUSY}}}}}
\newc{\mbmbmsbar}{\ensuremath{ m_b(m_b)^{\msbar} }}
\newc{\mbmbsmmsbar}{\ensuremath{ m_b(m_b)^{\msbar}_{{\mathrm{SM}}}}}
\newc{\mbmzsmmsbar}{\ensuremath{ m_b(\mz)^{\msbar}_{{\mathrm{SM}}}}}
\newc{\mbmzsmdrbar}{\ensuremath{ m_b(\mz)^{\drbar}_{{\mathrm{SM}}}}}
\newc{\mbmzmssmdrbar}{\ensuremath{ m_b(\mz)^{\drbar}_{{\mathrm{SUSY}}}}}
\newc{\mtaumzsmmsbar}{\ensuremath{ m_{\tau}(\mz)^{\msbar}_{{\mathrm{SM}}}}}
\newc{\mtaumzsmdrbar}{\ensuremath{ m_{\tau}(\mz)^{\drbar}_{{\mathrm{SM}}}}}
\newc{\mtaumzmssmdrbar}{\ensuremath{ m_{\tau}(\mz)^{\drbar}_{{\mathrm{SUSY}}}}}
\newc{\alphasmzms}{\ensuremath{\alpha_s(M_Z)^{\overline{MS}}}}
\newc{\alphaimzms}[1]{\ensuremath{\alpha_{#1}(M_Z)^{\overline{MS}}}}
\newc{\alphaemmz}{\ensuremath{\alpha_{\mathrm{em}}(M_Z)^{\overline{MS}}}}

\newc{\mzero}{\ensuremath{{m_0}}}
\newc{\mhalf}{\ensuremath{ m_{1/2}}}
\newc{\tanb}{\ensuremath{\tan\beta}}
\newc{\azero}{\ensuremath{ A_0}}
\newc{\bzero}{\ensuremath{ B_0}}
\newc{\signmu}{\ensuremath{\rm{sgn}\,\mu}}
\newc{\mueff}{\ensuremath{\mu_{\rm{eff}}}}
\newc{\lam}{\ensuremath{{\lambda}}}
\newc{\kap}{\ensuremath{{\kappa}}}
\newc{\alam}{\ensuremath{{A_{\lambda}}}}
\newc{\akap}{\ensuremath{{A_{\kappa}}}}
\newc{\hs}{\ensuremath{ H_s}}      
\newc{\mhs}{\ensuremath{ m_{H_s}}} 
\newc{\mgut}{\ensuremath{ M_{\rm GUT}}}
\newc{\mplanck}{\ensuremath{ M_{\rm P}}}      \newc{\mpl}{\ensuremath{ M_{\rm Pl}}}
\newc{\msusy}{\ensuremath{ M_{\rm SUSY}}}      \newc{\ms}{\ensuremath{ M_{\rm S}}}
 \newc{\mhl}{\ensuremath{m_\hl}} 
 \newc{\mhone}{\ensuremath{m_{h_1}}} 
 \newc{\mhtwo}{\ensuremath{m_{h_2}}} 
 \newc{\mglu}{\ensuremath{m_{\tilde g}}} 
 \newc{\mul}{\ensuremath{m_{\tilde{u}_L}}} 
 \newc{\mtone}{\ensuremath{m_{\tilde{t}_1}}} 
 \newc{\ma}{\ensuremath{m_A}} 
 \newc{\maone}{\ensuremath{m_{a_1}}} 
 \newc{\matwo}{\ensuremath{m_{a_2}}}
 \newc{\hone}{\ensuremath{h_1}}
 \newc{\htwo}{\ensuremath{h_2}}
 \newc{\aone}{\ensuremath{a_1}}
 \newc{\atwo}{\ensuremath{a_2}}
 \newc{\mhu}{\ensuremath{ m_{H_u}}}       
 \newc{\mhd}{\ensuremath{ m_{H_d}}}
 \newc{\mhusq}{\ensuremath{ m_{H_u}^2}}       
 \newc{\mhdsq}{\ensuremath{ m_{H_d}^2}}
 \newc{\mhuew}{\ensuremath{ m^{\ast}_{H_u}}}       
 \newc{\mhdew}{\ensuremath{ m^{\ast}_{H_d}}}
 \newc{\mhuewsq}{\ensuremath{ m^{\ast\, 2}_{H_u}}}       
 \newc{\mhdewsq}{\ensuremath{ m^{\ast\, 2}_{H_d}}}
 \newc{\hu}{\ensuremath{ H_u}}       
 \newc{\hd}{\ensuremath{ H_d}}
 \newc{\barmhu}{\ensuremath{ \bar{m}_{H_u}}}
 \newc{\barmhd}{\ensuremath{ \bar{m}_{H_d}}}

 \newc{\mqthree}{\ensuremath{m_{\widetilde{Q}_3}^2}}
 \newc{\muthree}{\ensuremath{m_{\tilde{u}_3}^2}}
 \newc{\mdthree}{\ensuremath{m_{\tilde{d}_3}^2}}
 \newc{\mlthree}{\ensuremath{m_{\widetilde{L}_3}^2}}
 \newc{\methree}{\ensuremath{m_{\tilde{e}_3}^2}}
 \newc{\mqtwo}{\ensuremath{m_{\widetilde{Q}_2}^2}}
 \newc{\mutwo}{\ensuremath{m_{\tilde{u}_2}^2}}
 \newc{\mdtwo}{\ensuremath{m_{\tilde{d}_2}^2}}
 \newc{\mltwo}{\ensuremath{m_{\widetilde{L}_2}^2}}
 \newc{\metwo}{\ensuremath{m_{\tilde{e}_2}^2}}
 \newc{\mqone}{\ensuremath{m_{\widetilde{Q}_1}^2}}
 \newc{\muone}{\ensuremath{m_{\tilde{u}_1}^2}}
 \newc{\mdone}{\ensuremath{m_{\tilde{d}_1}^2}}
 \newc{\mlone}{\ensuremath{m_{\widetilde{L}_1}^2}}
 \newc{\meone}{\ensuremath{m_{\tilde{e}_1}^2}}
 \newc{\msmul}{\ensuremath{m_{\tilde{\mu}_L}}}
 \newc{\msmur}{\ensuremath{m_{\tilde{\mu}_R}}}
 \newc{\msneumu}{\ensuremath{m_{\tilde{\nu}_{\mu}}}}
 \newc{\mone}{\ensuremath{M_1}}
 \newc{\monesq}{\ensuremath{M_1^2}}
 \newc{\mtwo}{\ensuremath{M_2}}
 \newc{\mtwosq}{\ensuremath{M_2^2}}
 \newc{\mthree}{\ensuremath{M_3}}
 \newc{\mthreesq}{\ensuremath{M_3^2}}
 \newc{\atau}{\ensuremath{{A_{\tau}}}}
 \newc{\at}{\ensuremath{{A_{t}}}}
 \newc{\ab}{\ensuremath{{A_{b}}}}
 \newc{\atausq}{\ensuremath{{A_{\tau}^2}}}
 \newc{\atsq}{\ensuremath{{A_{t}^2}}}
 \newc{\absq}{\ensuremath{{A_{b}^2}}}

 \newc{\dmzero}{\ensuremath{\Delta{_{m_0}}}}
 \newc{\dmhalf}{\ensuremath{\Delta{_{m_{1/2}}}}}
 \newc{\dmu}{\ensuremath{\Delta{_{\mu}}}}

 \newc{\pten}{\ensuremath{\psi_{10}}}
 \newc{\ffive}{\ensuremath{\phi_{5}}}
 \newc{\hfive}{\ensuremath{h_{5}}}
 \newc{\hbfive}{\ensuremath{h_{\bar{5}}}}
 \newc{\thet}{\ensuremath{\theta_{50}}}
 \newc{\thetb}{\ensuremath{\theta_{\,\overline{50}}}}
 \newc{\ptenhat}{\ensuremath{\hat{\psi}_{10}}}
 \newc{\ffivehat}{\ensuremath{\hat{\phi}_{5}}}
 \newc{\hfivehat}{\ensuremath{\hat{h}_{5}}}
 \newc{\hbfivehat}{\ensuremath{\hat{h}_{\bar{5}}}}
 \newc{\thethat}{\ensuremath{\hat{\theta}_{50}}}
 \newc{\thetbhat}{\ensuremath{\hat{\theta}_{\,\overline{50}}}}
 \newc{\si}{\ensuremath{\Sigma}}
 \newc{\mfive}{\ensuremath{m_5^2}}
 \newc{\mten}{\ensuremath{m_{10}^2}}
 \newc{\dfive}{\ensuremath{\Delta^2_5}}
 \newc{\dbfive}{\ensuremath{\Delta^2_{\bar{5}}}}
 \newc{\dfifty}{\ensuremath{\Delta^2_{50}}}
 \newc{\dfiftyb}{\ensuremath{\Delta^2_{\,\overline{50}}}}
 \newc{\msi}{\ensuremath{m_{\Sigma}^2}}
 \newc{\lamh}{\ensuremath{\lambda_{H}}}
 \newc{\lamhb}{\ensuremath{\lambda_{\bar{H}}}}
 \newc{\ah}{\ensuremath{A_{H}}}
 \newc{\ahb}{\ensuremath{A_{\bar{H}}}}
 \newc{\lams}{\ensuremath{\lambda_{S}}}
 \newc{\as}{\ensuremath{A_{S}}}
 \newc{\lamsig}{\ensuremath{\lambda_{\si}}}
 \newc{\asig}{\ensuremath{A_{\si}}}

 \newc{\msten}{\ensuremath{m_{16}^2}}
 \newc{\mhun}{\ensuremath{m_{126}^2}}
 \newc{\mhunb}{\ensuremath{m_{\bar{126}}^2}}
 \newc{\mthun}{\ensuremath{m_{210}^2}}
 \newc{\ahun}{\ensuremath{A_{\bar{126}}}}
 \newc{\yhun}{\ensuremath{Y_{\bar{126}}}}
 \newc{\aten}{\ensuremath{A_{10}}}
 \newc{\yten}{\ensuremath{Y_{10}}}
 \newc{\alone}{\ensuremath{A_{\lambda_1}}}
 \newc{\altwo}{\ensuremath{A_{\lambda_2}}}
 \newc{\althree}{\ensuremath{A_{\lambda_3}}}
 \newc{\althreeb}{\ensuremath{A_{\bar{\lambda_3}}}}
 \newc{\lone}{\ensuremath{\lambda_1}}
 \newc{\ltwo}{\ensuremath{\lambda_2}}
 \newc{\lthree}{\ensuremath{\lambda_3}}
 \newc{\lthreeb}{\ensuremath{\bar{\lambda_3}}}

\newc{\sigsip}{\ensuremath{\sigma^{\rm SI}_{p}}}	\newc{\sigsin}{\ensuremath{\sigma^{\rm SI}_{n}}}
\newc{\sigsdp}{\ensuremath{\sigma^{\rm SD}_{p}}}	\newc{\sigsdn}{\ensuremath{\sigma^{\rm SD}_{n}}}
\newc{\sigsi}{\ensuremath{\sigma^{\rm SI}}}	\newc{\sigsd}{\ensuremath{\sigma^{\rm SD}}}
\newc{\sigv}{\ensuremath{\sigma v}}
\newc{\abund}{\ensuremath{ \Omega h^2}}
\newc{\omegadm}{\ensuremath{ \Omega_{{\rm DM}}}}     \newc{\abunddm}{\ensuremath{ \Omega_{{\rm DM}} h^2}} 
\newc{\omegam}{\ensuremath{ \Omega_{{\rm m}}}}       \newc{\abundm}{\ensuremath{ \Omega_{{\rm m}} h^2}}
\newc{\omegab}{\ensuremath{ \Omega_{{\rm b}}}}	\newc{\abundb}{\ensuremath{ \Omega_{{\rm b}} h^2}}
\newc{\omegatot}{\ensuremath{ \Omega_{{\rm TOT}}}}
\newc{\omegacdm}{\ensuremath{ \Omega_{{\rm CDM}}}}   \newc{\abundcdm}{\ensuremath{ \Omega_{{\rm CDM}} h^2}}
\newc{\omegalambda}{\ensuremath{ \Omega_{\Lambda}}} \newc{\abundlambda}{\ensuremath{ \Omega_{\Lambda} h^2}}
\newc{\omegarad}{\ensuremath{ \Omega_{{\rm rad}}}}  \newc{\abundrad}{\ensuremath{ \Omega_{{\rm rad}} h^2}}
\newc{\rhocrit}{\ensuremath{ \rho_{\rm crit}}}
\newc{\rhochi}{\ensuremath{ \rho_{\chi}}}
\newc{\abunchi}{\ensuremath{\Omega_\chi h^2}}
\newc{\abundlsp}{\ensuremath{\Omega_{\rm LSP}h^2}}

\newc{\amu}{\ensuremath{ a_{\mu}}}        \newc{\amususy}{\ensuremath{ a_{\mu}^{\mathrm{SUSY}}}}
\newc{\amuexpt}{\ensuremath{ a_{\mu}^{\mathrm{expt}}}}        \newc{\amusm}{\ensuremath{ a_{\mu}^{\mathrm{SM}}}}
\newc\deltaamu{\ensuremath{\Delta a_{\mu}}} \newc{\deltaamususy}{\ensuremath{\delta a_{\mu}^{\mathrm{SUSY}}}}
\newc\gmtwo{\ensuremath{ (g-2)_{\mu}}} 
\newc{\deltagmtwomususy}{\ensuremath{\delta\left(g-2\right)_{\mu}^{\mathrm{SUSY}}}}
\newc{\deltagmtwomu}{\ensuremath{\delta\left(g-2\right)_{\mu}}}
\newc\BR{\ensuremath{\textrm{BR}}}
\newc\bsgamma{\ensuremath{ b\rightarrow s \gamma }}
\newc\bxsgamma{\ensuremath{\overline{B}\rightarrow X_{s}\gamma}}
\newc\brbsgamma{\ensuremath{\BR\left(\bsgamma\right)}}
\newc\brbxsgamma{\ensuremath{\BR\left(\bxsgamma\right)}}
\newc\bsmumu{\ensuremath{B_s\to\mu^+\mu^-}}
\newc\brbsmumu{\ensuremath{\BR\left(B_s\to\mu^+\mu^-\right)}}
\newc\bdmmumu{\ensuremath{\overline{B}_d\to\mu^+\mu^-}}
\newc\bbbarmix{\ensuremath{\overline{B}_s\mbox{-}B_s}}      
\newc\delmbs{\ensuremath{\Delta M_{B_s}}}
\newc{\butaunu}{\ensuremath{B_u \rightarrow \tau \nu}}
\newc{\brbutaunu}{\ensuremath{\BR\left(B_u \rightarrow \tau \nu\right)}}

\newc{\brmuegamma}{\ensuremath{\BR\left(\mu^{\pm}\rightarrow e^{\pm}\gamma\right)}}
\newc{\brtauegamma}{\ensuremath{\BR\left(\tau^{\pm}\rightarrow e^{\pm}\gamma\right)}}
\newc{\brtaumugamma}{\ensuremath{\BR\left(\tau^{\pm}\rightarrow \mu^{\pm}\gamma\right)}}
\newc{\brmuthreee}{\ensuremath{\BR\left(\mu^{\pm}\rightarrow e^{\pm}e^+e^-\right)}}
\newc{\brtauthreee}{\ensuremath{\BR\left(\tau^{\pm}\rightarrow e^{\pm}e^+e^-\right)}}
\newc{\brtauthreemu}{\ensuremath{\BR\left(\tau^{\pm}\rightarrow \mu^{\pm}\mu^+\mu^-\right)}}

\newcommand*{\charone}{\ensuremath{\chi^{\pm}_1}}

\newcommand*{\SARAH}{SARAH}

\let\oldcite\cite
\renewcommand*{\cite}{~\oldcite}

\newcommand*{\hl}{\ensuremath{h}}

\newcommand*{\mfivem}{\textbf{LD}}

\newc{\glzmu}{\ensuremath{{g^{Z\mu \mu}_{L}}}}
\newc{\grzmu}{\ensuremath{{g^{Z\mu \mu}_{R}}}}
\newc{\glwmu}{\ensuremath{{g^{W\mu \nu_\mu}_{L}}}}
\newc{\grwmu}{\ensuremath{{g^{W\mu \nu_\mu}_{R}}}}
\newc{\glzmuSM}{\ensuremath{{g^{Z\mu \mu}_{L,\textrm{SM}}}}}
\newc{\grzmuSM}{\ensuremath{{g^{Z\mu \mu}_{R,\textrm{SM}}}}}
\newc{\glwmuSM}{\ensuremath{{g^{W\mu \nu_\mu}_{L,\textrm{SM}}}}}
\newc{\grwmuSM}{\ensuremath{{g^{W\mu \nu_\mu}_{R,\textrm{SM}}}}}

\restylefloat{figure}

\def\beq {\begin{equation}}
\def\eeq {\end{equation}}
\def\bi {\begin{itemize}}
\def\ei {\end{itemize}}
\def\bea {\begin{eqnarray}}
\def\eea {\end{eqnarray}}

\def \met{\rm E{\!\!\!/}_T}

\newcommand{\br}{\begin{eqnarray}}
\newcommand{\er}{\end{eqnarray}}

\newcommand{\ch}{\widetilde \chi^\pm}

\def\lum             {{\cal L}}
\newcommand{\ifb} {\rm {fb}^{-1}}

\def\chisq{\chi^2}

\newc{\wt}{\widetilde}

\def \ch2p {{\wt\chi_2^+}}

\def \ch2m {{\wt\chi_2^-}}

\def \chonepm{{\chi_{1}^{\pm}}}
\def \chonemp{{\wt\chi_1}^{\mp}}
\def \mchonepm{m_{\chonepm}}
\def \mchonemp{m_{\chonemp}}

\newc{\dmchi}{\Delta m_{\wt\chi}}

\def \lspone{\chi_1^0}
\def \mlspone{m_{\lspone}}
\def \lsptwo{\chi_2^0}
\def \mlsptwo{m_{\lsptwo}}

\def \slepton{\wt l}

\def\issue(#1,#2,#3){{\bf #1}, #2 (#3)}
\def\iss(#1,#2,#3){{\bf #1} (#3) #2}
\def\ASTR(#1,#2,#3){Astropart.\ Phys. \issue(#1,#2,#3)}
\def\AJ(#1,#2,#3){Astrophysical.\ Jour. \issue(#1,#2,#3)}
\def\AJS(#1,#2,#3){Astrophys.\ J.\ Suppl. \issue(#1,#2,#3)}
\def\APP(#1,#2,#3){Acta.\ Phys.\ Pol. \issue(#1,#2,#3)}
\def\JCAP(#1,#2,#3){Journal\ XX. \issue(#1,#2,#3)} 
\def\SC(#1,#2,#3){Science \issue(#1,#2,#3)}
\def\PRD(#1,#2,#3){Phys.\ Rev.\ D \issue(#1,#2,#3)}
\def\PR(#1,#2,#3){Phys.\ Rev.\ \issue(#1,#2,#3)} 
\def\PRC(#1,#2,#3){Phys.\ Rev.\ C \issue(#1,#2,#3)}
\def\NPB(#1,#2,#3){Nucl.\ Phys.\ B \issue(#1,#2,#3)}
\def\NPPS(#1,#2,#3){Nucl.\ Phys.\ Proc. \ Suppl \issue(#1,#2,#3)}
\def\NJP(#1,#2,#3){New.\ J.\ Phys. \issue(#1,#2,#3)}
\def\JP(#1,#2,#3){J.\ Phys.\issue(#1,#2,#3)}
\def\PL(#1,#2,#3){Phys.\ Lett. \issue(#1,#2,#3)}
\def\ZP(#1,#2,#3){Z.\ Phys. \issue(#1,#2,#3)}
\def\ZPC(#1,#2,#3){Z.\ Phys.\ C  \issue(#1,#2,#3)}
\def\PREP(#1,#2,#3){Phys.\ Rep. \issue(#1,#2,#3)}
\def\PRL(#1,#2,#3){Phys.\ Rev.\ Lett. \issue(#1,#2,#3)}
\def\MPL(#1,#2,#3){Mod.\ Phys.\ Lett. \issue(#1,#2,#3)}
\def\RMP(#1,#2,#3){Rev.\ Mod.\ Phys. \issue(#1,#2,#3)}
\def\SJNP(#1,#2,#3){Sov.\ J.\ Nucl.\ Phys. \issue(#1,#2,#3)}
\def\CPC(#1,#2,#3){Comp.\ Phys.\ Comm. \issue(#1,#2,#3)}
\def\IJMPA(#1,#2,#3){Int.\ J.\ Mod. \ Phys.\ A \issue(#1,#2,#3)}
\def\MPLA(#1,#2,#3){Mod.\ Phys.\ Lett.\ A \issue(#1,#2,#3)}
\def\PTP(#1,#2,#3){Prog.\ Theor.\ Phys. \issue(#1,#2,#3)}
\def\RMP(#1,#2,#3){Rev.\ Mod.\ Phys. \issue(#1,#2,#3)}
\def\NIMA(#1,#2,#3){Nucl.\ Instrum.\ Methods \ A \issue(#1,#2,#3)}
\def\EPJC(#1,#2,#3){Eur.\ Phys.\ J.\ C \issue(#1,#2,#3)}
\def\RPP (#1,#2,#3){Rept.\ Prog.\ Phys. \issue(#1,#2,#3)}
\def\PPNP(#1,#2,#3){ Prog.\ Part.\ Nucl.\ Phys. \issue(#1,#2,#3)}
\newc{\PRDR}[3]{{Phys. Rev. D} {\bf #1}, Rapid  Communications, #2 (#3)}

\def\PLB(#1,#2,#3){Phys.\ Lett.\ B  \iss(#1,#2,#3)}
\def\JHEP(#1,#2,#3){JHEP \iss(#1,#2,#3)}

\title{Impact of LHC data on muon {\boldmath $g-2$} solutions in
  vector-like extension of the Constrained MSSM }

\author[a]{Arghya Choudhury,}
\author[b]{Soumya Rao,}
\author[a,b]{and Leszek Roszkowski}

\affiliation{$^a$ Consortium for Fundamental Physics, Department of Physics and Astronomy,\\ University of Sheffield, Sheffield S3 7RH, United Kingdom\\ 
 $^b$ National Centre for Nuclear Research,\\ Ho{\.z}a 69, 00-681 Warsaw, Poland} 

\emailAdd{a.choudhury@sheffield.ac.uk}
\emailAdd{soumya.rao@ncbj.gov.pl}
\emailAdd{leszek.roszkowski@ncbj.gov.pl}

\abstract{The long--standing discrepancy between the experimental
  determination by the Muon $g-2$ Collaboration at Brookhaven and the
  Standard Model predictions for the anomalous magnetic moment of the
  muon cannot be explained within simple unified framework like the
  Constrained Minimal Supersymmetric Standard Model, but it can within
  its extension with vector-like fermions. In this paper we consider a
  model with an additional vector-like $5+\bar{5}$ pair of
  $SU(5)$. Within this model we first identify its parameter
  space that is consistent with the current discrepancy and show that
  this implies the lighter chargino mass in the range of $700-1200$
  GeV. We examine how it is affected by constraints from electroweak
  sparticle search at the LHC based on 13 TeV search with $36.1\ifb$
  integrated luminosity. We show that null trilepton signal searches
  coming from chargino--neutralino pair production significantly
  constrains the allowed parameter space except when the
  chargino--neutralino mass difference is relatively small, below about
  10 GeV.  Next we consider the expected impact of the New Muon $g-2$
  experiment at Fermilab with its projected sensitivity reach of
  $7\,\sigma$ and, assuming it confirms the current discrepancy, show
  that the remaining parameter space of the considered model will be
  in strong tension with the current LHC limits.}

\begin{document}
 \maketitle

\section{Introduction}

Despite the absence of a signal for supersymmetry (SUSY) at the LHC, it still remains one
of the most appealing frameworks for physics beyond the Standard Model (BSM).  Besides
providing a natural candidate for dark matter (DM), it also gives possible explanation for
the discrepancy that exists in the Standard Model (SM) value of the anomalous magnetic
moment of muon, \gmtwo, and the experimentally measured quantity.  The SM value for
\gmtwo\ differs by more than $3\sigma$ from the measured
value\cite{Bennett:2006fi,Miller:2007kk}.  Future measurement at
Fermilab\cite{Grange:2015fou,Chapelain:2017syu} is expected to improve the sensitivity of
the previous measurement by a factor of four and hence potentially confirm or falsify the
persistent disagreement.  In SUSY, the explanation for the difference arises from
contributions due to smuon-neutralino and sneutrino-chargino loops.  To fit the  \gmtwo
~anomaly within the framework  of the minimal supersymmetric Standard Model (MSSM), one
requires the slepton and the lighter chargino masses in a range of a few hundreds of
GeV\cite{Endo:2013bba,Fowlie:2013oua,Chakraborti:2014gea,Das:2014kwa,Chakraborti:2015mra,Lindner:2016bgg,Endo:2017zrj}.
However, the stringent bounds on the strong sector (squarks and gluinos) from the LHC and
the  Higgs mass measurements rule out the possibility of explaining \gmtwo~  in
GUT-constrained models like the Constrained MSSM (CMSSM) and the Non-Universal Higgs Mass
(NUHM) model\cite{Fowlie:2012im, Bechtle:2015nua,Cao:2011sn}.  The way out has usually been to assume
non-universal gaugino
masses\cite{Mohanty:2013soa,Akula:2013ioa,Chakrabortty:2013voa,Kowalska:2015zja,
Chakrabortty:2015ika,Belyaev:2016oxy,Okada:2016wlm,Fukuyama:2016mqb}
which can provide a viable parameter space to explain \gmtwo ~while at the same time not
conflicting with constraints from LEP and LHC.

There have also been alternative solutions as for example, adding vector-like (VL) matter
to MSSM which has been studied in
Refs.\cite{Endo:2011xq,Endo:2011mc,Dermisek:2013gta,Dermisek:2014cia,
Gogoladze:2015jua,Aboubrahim:2016xuz,Nishida:2016lyk,Higaki:2016yeh,Megias:2017dzd,
Choudhury:2017fuu}.  The presence of new VL sector gives extra contribution to \gmtwo ~by
introducing new sources of smuon mixing and new Yukawa couplings\cite{Choudhury:2017fuu}.
Apart from \gmtwo, it has been shown that VL colored sparticles provide extra
contributions to Higgs
mass\cite{Graham:2009gy,Martin:2009bg,Faroughy:2014oka,Lalak:2015xea,Nickel:2015dna,Barbieri:2016cnt}
and several phenomenological analyses have addressed the extra VL matter in the context of
various long-standing theoretical issues related to beyond SM
physics\cite{Endo:2011xq,Endo:2011mc,Dermisek:2013gta,Dermisek:2014cia,
Gogoladze:2015jua,Aboubrahim:2016xuz,Nishida:2016lyk,Higaki:2016yeh,Megias:2017dzd,
Choudhury:2017fuu}. 

In particular, Ref.\cite{Choudhury:2017fuu} studied two simple extensions of the CMSSM by
adding a pair of multiplets, firstly in the $\mathbf{5}+\mathbf{\bar{5}}$ and secondly in
the $\mathbf{10}+\mathbf{\overline{10}}$ representations of $SU(5)$.  It was shown that
the model could satisfy various constraints from flavor physics and LHC direct searches,
as well as include a viable dark matter candidate that was in agreement with relic density
and direct detection constraints, for a considerable region of the parameter
space\cite{Choudhury:2017fuu}.  In particular, through the additional mixing of VL fields
with second generation leptons, the model proved particularly useful in explaining the
\gmtwo\ discrepancy.  In this work we extend the analysis considered in
Ref.\cite{Choudhury:2017fuu}, using the models with additional
$\mathbf{5}+\mathbf{\bar{5}}$.  Motivated by the solution to the \gmtwo\ discrepancy, we
examine the impact of collider constraints on the viable parameter space.

As mentioned earlier, the allowed parameter spaces satisfying \gmtwo\ constraints are
characterised by light electroweak (EW) sparticles, i.e.,  light EW gauginos or
electroweakinos (the charginos and the neutralinos) and charged sleptons. Hence to probe
the relevant parameter space at the LHC, the most sensitive search is chargino and
neutralino pair production (via $pp \rightarrow \chonepm \lsptwo$) leading to the
trilepton + transverse missing energy ($\met$) signal. Both CMS and ATLAS Collaborations
have looked for electroweakinos, or EWinos, with different leptonic final states\cite{Aad:2014nua,Aad:2014vma,Aad:2014yka,Aad:2015jqa,Aad:2015eda,
Khachatryan:2014mma,Khachatryan:2014qwa,atlas_ew_3l,CMS-PAS-SUS-17-004}, among which the
trilepton data gives the most stringent bound. From the very recent LHC analysis of Run-II
data with $\lum$ = 36.1 $\ifb$, ATLAS has excluded chargino masses upto 1150 GeV for
relatively light LSP\cite{atlas_ew_3l}. However, ATLAS and CMS have presented these limits
for a few particular type of simplified models with specific assumptions on the
compositions and branching ratios of EWinos.  The electroweakinos searches and related
topics in the context of the LHC have been analysed by various phenomenology group in
Ref.\cite{Bharucha:2013epa, Howe:2012xe, Schwaller:2013baa, Ghosh:2012mc, Eckel:2014dza,
	Choudhury:2013jpa, Han:2014xoa, Yu:2014mda, Papaefstathiou:2014oja, Ajaib:2015yma,
Han:2013kza, Choudhury:2016lku,  Xiang:2016ndq, Datta:2016ypd, Hagiwara:2017lse,
Arbey:2017eos, Chakraborti:2017vxz, Chakraborti:2014gea, Chakraborti:2015mra,Cao:2015efs,Kobakhidze:2016mdx}.  Due to the
presence of VL particles and their mixing with SM, the electroweakinos (mainly $\chonepm,
\lsptwo$) can have non standard branching ratios compared to the CMSSM or usual
phenomenological MSSM (pMSSM) scenarios. Hence the limits interpreted by ATLAS or CMS for
various simplified models are not directly applicable to such models and a
reinterpretation of the bounds from trilepton searches at the LHC is necessary.

The paper is organized as follows. We first give a brief overview of the model which is
obtained by adding a VL  $\mathbf{5}+\mathbf{\bar{5}}$ of $SU(5)$ pair to CMSSM in
Section~\ref{sec:model}.  We briefly mention the constraints applied to obtain the
relevant allowed parameter space in Section~\ref{constraints} and then discuss different
scenarios for chargino ($\chonepm$) and neutralino ($\lsptwo$) decays in the context of VL
extension of CMSSM in Section~\ref{sec:branchings}.  In Section~\ref{trilepton}, we show
results for LHC trilepton searches from  chargino-neutralino pair production using the
latest LHC Run-II $36.1 fb^{-1}$ data. Finally, we give our conclusions in
Section~\ref{conclusions}.

\section{Vector like extension of the CMSSM\label{sec:model}}

We follow the model studied and analysed in\cite{Choudhury:2017fuu}, particularly in the
context of \gmtwo\ where  the MSSM is extended through the addition of a pair
$\mathbf{5}+\mathbf{\bar{5}}$ or a pair $\mathbf{10}+\mathbf{\overline{10}}$.  However, it
was shown in \cite{Choudhury:2017fuu} that the $\mathbf{10}+\mathbf{\overline{10}}$
extension was more fine tuned in order to provide a viable parameter space for a
significant contribution to \gmtwo\ and therefore the analysis was restricted to
$\mathbf{5}+\mathbf{\bar{5}}$.  Here we shall focus only on the
$\mathbf{5}+\mathbf{\bar{5}}$ extension, which we shall from now on refer to as the LD
model following the previous convention.  We summarize the main features of the LD model
below (for more details see Ref.\cite{Choudhury:2017fuu}).


The LD model consists of extending the MSSM spectrum with the addition of the following
fields:\footnote{The MSSM fields are
$q$=(3,2,1/6), $u$=($\bar{3}$,1,-2/3), $d$=($\bar{3}$,${1}$,1/3), 
$l$=($\bf{1}$,$\bf{2}$,-1/2),   $e$=($\bf{1}$,$\bf{2}$,-1/2), 
$H_u$=(1,2,1/2),   $H_d$=(1,2,-1/2) with $SU(3)\times SU(2)\times U(1)$ 
quantum numbers in parentheses. }
\begin{align*}
D&=(\mathbf{\bar{3}},\mathbf{1},1/3)& D'&=(\mathbf{3},\mathbf{1},-1/3)\nonumber\\
L&=(\mathbf{1},\mathbf{2},-1/2)& L'&=(\mathbf{1},\mathbf{2},1/2)\,.
\end{align*}

This implies the addition of a quark with charge $-1/3$ and a charged lepton along with
their antiparticles, and two massive neutrinos to the MSSM spectrum.  Correspondingly the
sparticle content sees the addition of squarks, sleptons and sneutrinos.

In comparison to the MSSM, there are now additional trilinear and bilinear terms in the
superpotential,
\be
W \supset -\lam_D\,q H_d D-\lam_L\,L H_d e+M_D D D'+M_L L L'+\widetilde{M}_L\,lL'+\widetilde{M}_D\,dD'\,,\label{superpot5}
\ee

Finally the soft SUSY-breaking Lagrangian also has extra terms involving
$\tilde{L}^{(\prime)}$ and $\tilde{D}^{(\prime)}$ as follows
\bea
-\mathcal{L}_{\textrm{soft}}&\supset&\left[m_L^2|\tilde{L}|^2+m_{L'}^2|\tilde{L}'|^2
+m_D^2|\tilde{D}|^2+m_{D'}^2|\tilde{D}'|^2+\left(\widetilde{m}_L^2\,\tilde{l}^{\dag}\tilde{L}
+\widetilde{m}_D^2\,\tilde{d}^{\dag}\tilde{D}+\textrm{h.c.}\right)\right]\nonumber\\
 & & +\left(B_{M_L}\tilde{L}\tilde{L}'+B_{\widetilde{M}_L}\tilde{l}\tilde{L}'
+B_{M_D}\tilde{D}\tilde{D}'+B_{\widetilde{M}_D}\tilde{d}\tilde{D}'+\textrm{h.c.}\right) \nonumber \\ 
& &-\left(T_D\,\tilde{q}H_d\tilde{D}^{\dag}
+T_L\,\tilde{L}H_d\tilde{e}^{\dag}+\textrm{h.c.}\right),\label{soft5}
\eea
where $\widetilde{m}_L^2$, $\widetilde{m}_D^2$, $T_L$, $T_D$, $B_{\widetilde{M}_L}$, and
$B_{\widetilde{M}_D}$ are 3-dimensional matrices that govern mixing between MSSM and VL
matter.  This mixing plays an important role for \gmtwo\ phenomenology.

In addition to the above, we also make the choice of GUT scale parameters such that the
boundary conditions for the extra Yukawa couplings demanded by UV completion at the GUT
scale are given by
\begin{equation}
	\lambda_L=\begin{pmatrix}
		  0\\
	  	  \lambda_5\\
	  	  \epsilon\lambda_5
	  \end{pmatrix},
	\label{bc}
\end{equation}
where $\epsilon<1$.
This in turn means that the soft mass matrices which also satisfy the same flavor
constraints as the Yukawa couplings, will have their off-diagonal mixing terms
parametrized similar to Eqn.~\ref{bc} as follows

\begin{equation}
	\wt{m}_L^2=\wt{m}_D^2=\begin{pmatrix}
		  0\\
		  \wt{m}^2\\
		  \alpha\wt{m}^2
	  \end{pmatrix},
	\label{offdiag}
\end{equation}
where once again $\alpha<1$.

Thus the first generation mixing is almost absent compared to second and third generation
mixing.  Eqns.~\ref{bc} and \ref{offdiag} not only impact the \gmtwo\ contribution but
also have a significant effect on the trilepton signal from chargino and neutralino decays
as we shall see in Section~\ref{sec:branchings}.


\section{Constraints from flavor physics, \gmtwo\ and direct
	detection of DM\label{constraints}}

In this section we mention the GUT scale input parameters used as well as the constraints
applied in order to obtain the parameter space shown in
Fig.~\ref{fig:gm2VL5scan}\footnote{The result shown in Fig.~\ref{fig:gm2VL5scan} is
obtained using the same numerical tools and priors used in Ref.\cite{Choudhury:2017fuu}.}.
The parameter space was scanned using \texttt{MultiNest}\cite{Feroz:2008xx} and  the
\SARAH \cite{staub_sarah_2014,staub_automatic_2011,porod:2011nf,Staub:2011dp} package was used to generate the
spectrum, while the
relevant flavor constraints were calculated using the \SARAH\ package
FlavorKit\cite{Porod:2014xia}. In addition, dark matter constraints on relic density and
direct detection were obtained using $\tt micrOMEGAs\ v.3.5.5$\cite{Belanger:2013oya}.
Bounds on the Higgs sector from LHC searches for Higgs production channels, branching
ratios as well as Higgs decay were applied using the codes $\tt
HiggsSignals$\cite{Bechtle:2013xfa} and $\tt
HiggsBounds$\cite{Bechtle:2008jh,Bechtle:2011sb,Bechtle:2013wla}.  The following ranges of
values for the GUT scale input parameters were used, which are also listed
in\cite{Choudhury:2017fuu}:

\begin{align*}
	\mbox{VL Yukawa coupling, } \lambda_5 &\in[-0.5,0.5],\\
	\mbox{Yukawa hierarchy factor, } \epsilon &\in[-0.5,0.5],\\
	\mbox{superpotential mass VL fields, }M_V &\in[50,1500] \mbox{ GeV},\\
	\mbox{superpotential mass mixing, }\wt{M} &\in[-20,20] \mbox{ GeV},\\
	\mbox{mass mixing hierarchy factor, }\alpha &\in[0.01,1),\\
	\mbox{scalar mass, }m_0 &\in[100,4000] \mbox{ GeV},\\
	\mbox{gaugino mass, }m_{1/2} &\in[300,4000] \mbox{ GeV},\\
	\mbox{soft mass mixing, }\wt{m}^2 &\in[-5\times 10^6,5\times 10^6] \mbox{ GeV}^2,\\
	\mbox{trilinear coupling, }A_0 &\in[-4000,4000] \mbox{ GeV},\\
	\mbox{soft bilinear term VL fields, }B_0 &\in[-1500,1500] \mbox{ GeV},\\
	\mbox{ratio of the Higgs vevs, }\tan\beta &\in[1,60],
	\label{gutinput}
\end{align*}
and the sign of the Higgs mass parameter, sgn $\mu=+1$.

The experimental constraints used to derive the parameter space in addition to the Higgs
bounds are flavor physics constraints such as $\brbxsgamma$\cite{Amhis:2016xyh},
$\brbutaunu$\cite{Adachi:2012mm}, \delmbs\cite{Olive:2016xmw}, $\Delta
\rho$\cite{Olive:2016xmw}, $\brbsmumu$\cite{Aaij:2013aka,Chatrchyan:2013bka} and
$\brtaumugamma$\cite{Aubert:2009ag}, while in DM sector the constraint on relic
abundance\cite{Ade:2015xua}, \abunchi, and the current LUX limit\cite{Akerib:2016vxi} on
the spin-independent DM-nucleon scattering cross section, \sigsip, are taken into account.
For more details on ranges and theoretical and experimental errors see Table 1 of
ref.\cite{Choudhury:2017fuu}.

\section{Allowed parameter space and decay properties of EWinos \label{sec:branchings}}

\begin{figure}[!b]
\centering
\includegraphics[width=0.75\textwidth]{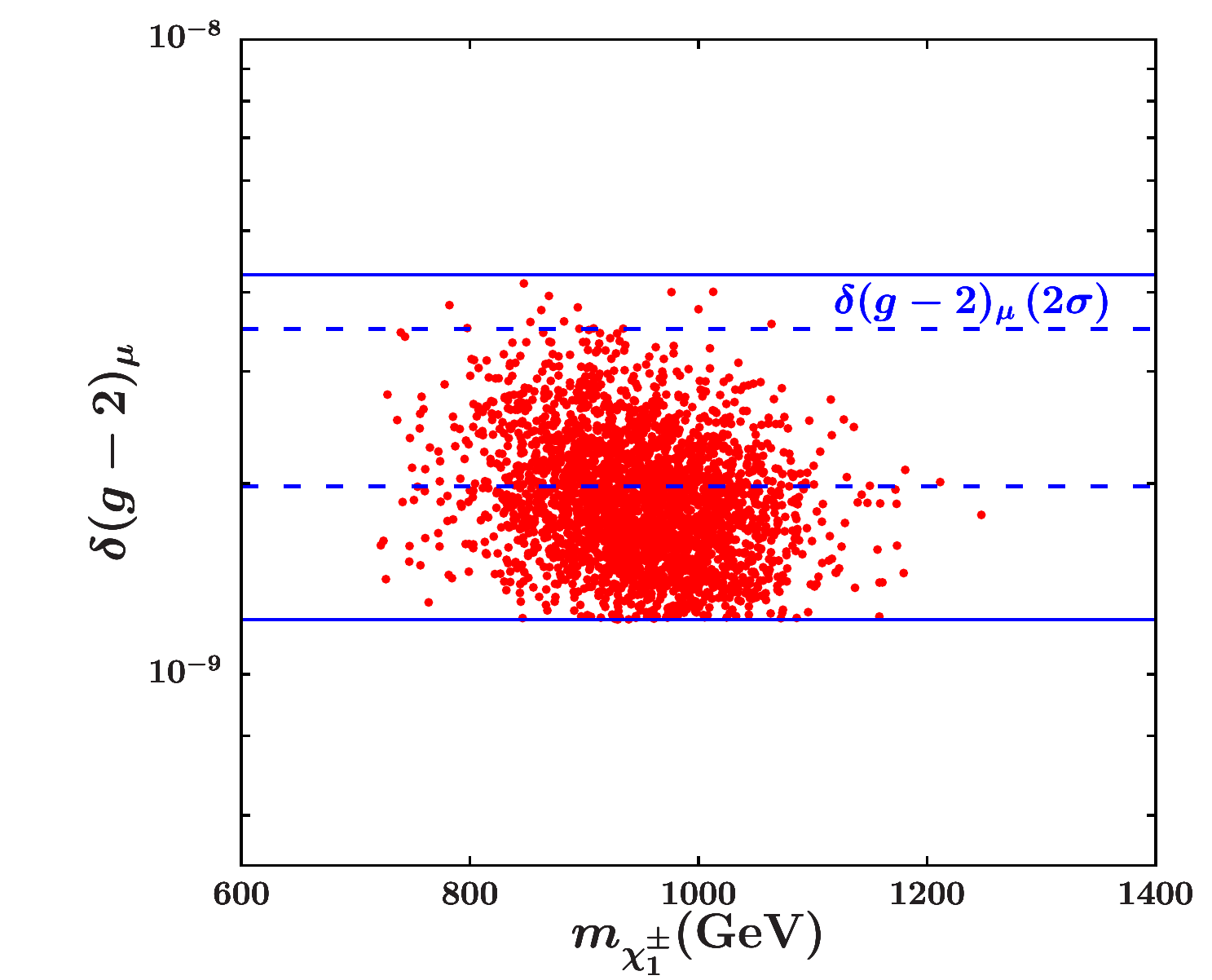}
\caption{Allowed parameter space for \deltagmtwomu\ in the \mfivem\ model as a function of
	chargino mass. The $2\,\sigma$ allowed region for \deltagmtwomu\ according to the
	latest data\cite{Bennett:2006fi,Miller:2007kk} is indicated by the blue solid
	lines, while the dashed lines indicate future
	measurement\cite{Grange:2015fou,Chapelain:2017syu} with four times greater
	sensitivity, assuming that the central value remains the same.}
\label{fig:gm2VL5scan}
\end{figure}

In this section we study decay properties of $\chonepm$ and $\lsptwo$ in the parameter
space which satisfies the constraints mentioned in the previous section as well as
\deltagmtwomu\ bounds ($2\,\sigma$) as shown in Fig.~\ref{fig:gm2VL5scan}.  The
$2\,\sigma$ allowed region for \deltagmtwomu\ according to the latest
data\cite{Bennett:2006fi,Miller:2007kk} is indicated by the blue solid lines, while the
dashed lines indicate future measurement\cite{Grange:2015fou,Chapelain:2017syu} with four
times greater sensitivity, assuming that the central value remains the same.  We consider
the red points that are allowed by current \deltagmtwomu\ bounds  for further analysis.
The trilepton final states from direct chargino-neutralino pair production will be the
most effective channel to probe all these points. Due to the choice of input parameters
and mixing in our model, it is expected that the decay modes of EWinos could be different
from the MSSM cases or usual choices made by ATLAS/CMS with simplified scenarios.

\begin{figure}[t]
\centering
\subfloat{%
\hspace{-2.1cm}
\includegraphics[width=0.6\textwidth]{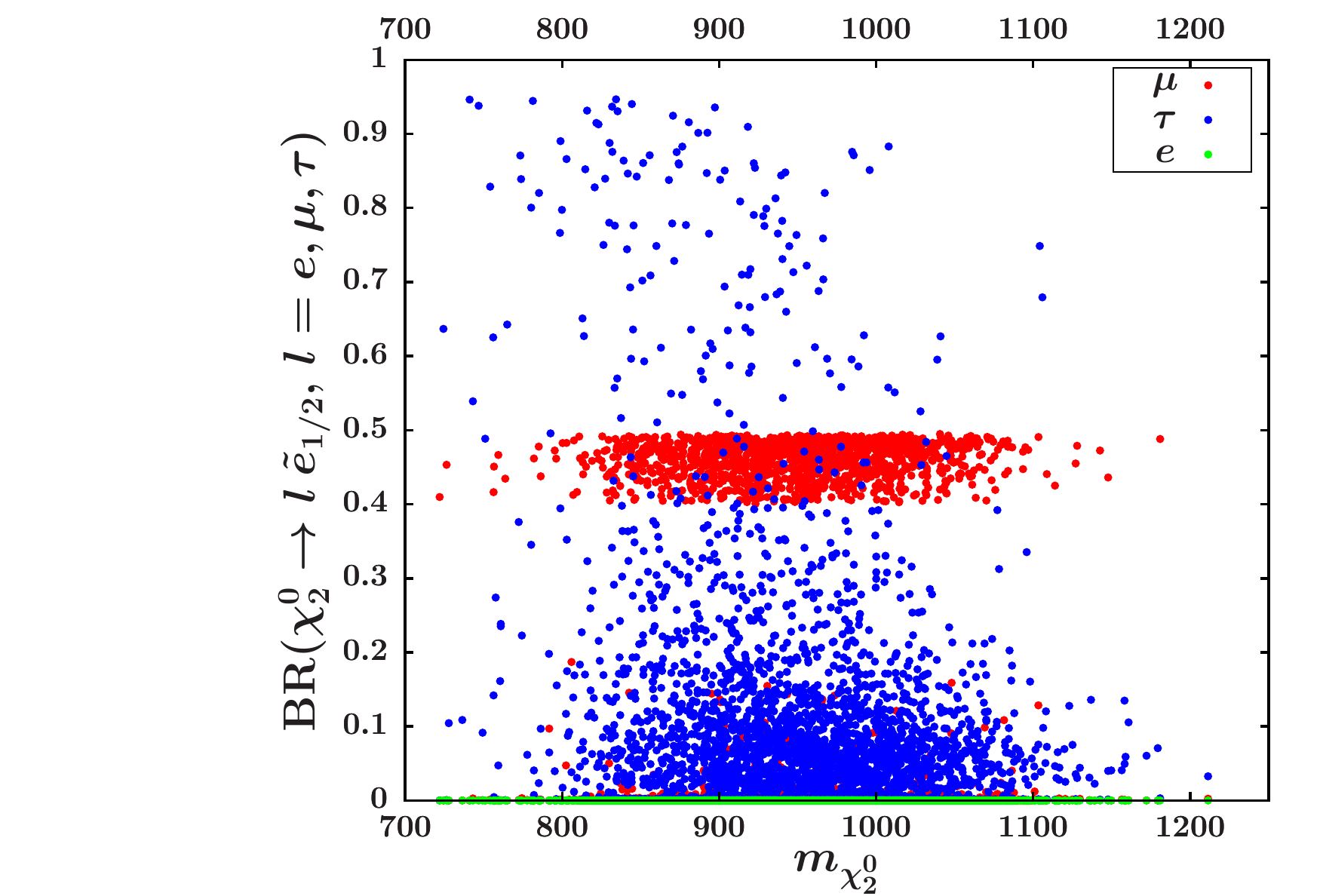}
}%
\subfloat{%
\hspace{-0.07\textwidth}
\includegraphics[width=0.6\textwidth]{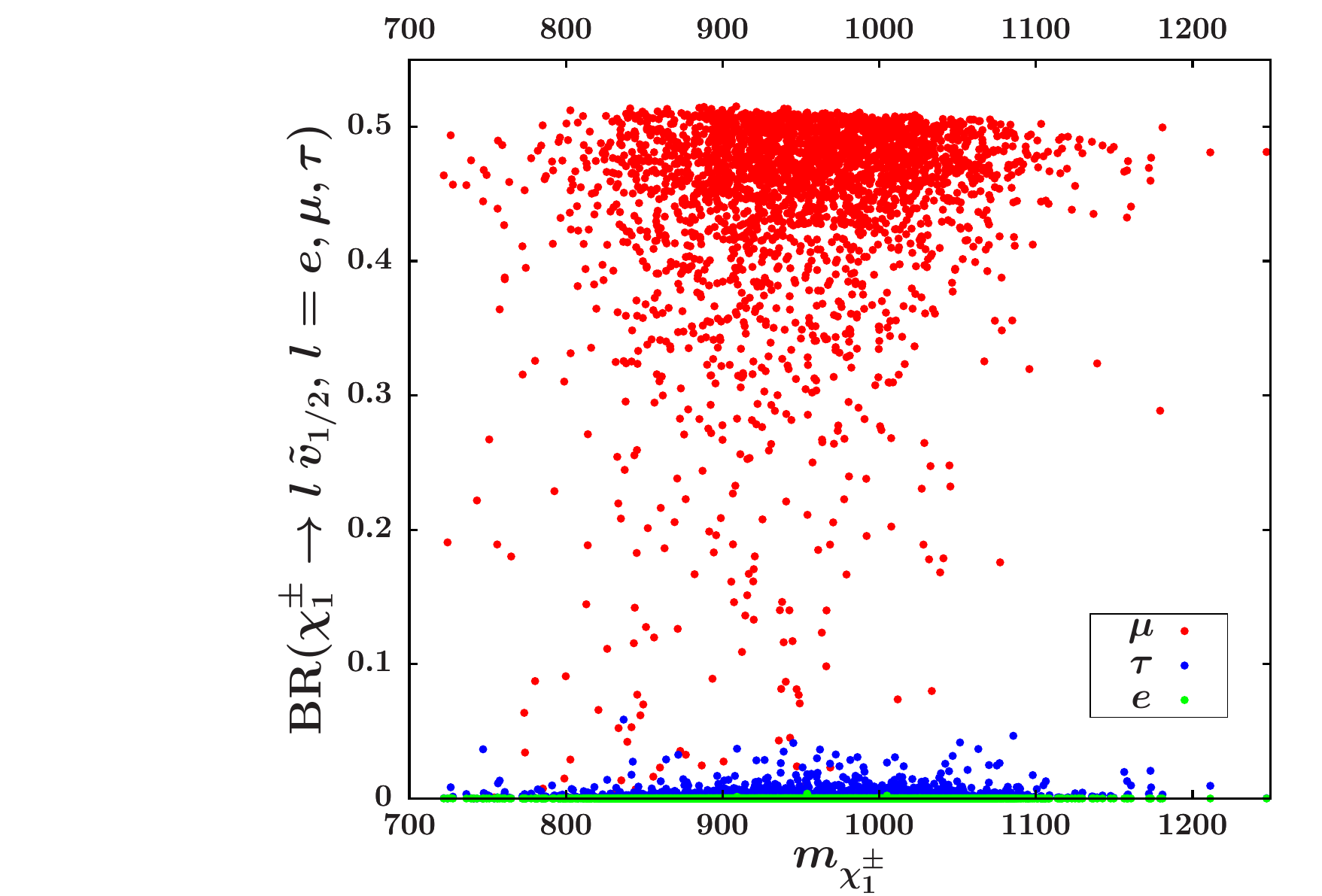}
}%
\caption{Branching ratio of $\lsptwo\to l\tilde{e}$ (left panel) and $\chonepm\to
l\tilde{\nu}$ (right panel) plotted against $\lsptwo$ and $\chonepm$ masses, respectively,
for the
parameter space satisfying the constraints described in the text as
well as giving a \gmtwo\  contribution that is within the current $2\sigma$ limit.}
\label{fig:decmode}
\end{figure}

\subsection{Decay modes for $\lsptwo$:}

In general for light slepton scenarios (lighter than $\mchonepm$), the second lightest
neutralino $\lsptwo$ decays into 2 body final state -- $l \tilde{l}$ and $\nu \tilde \nu$
where $l$ denotes for $e, \mu$ and $\tau$. For our model, the first slepton mass
eigenstate is mostly mixed smuon/VL and the second slepton eigenstate is usually
right-handed stau. Hence we sometimes  obtain large mass splitting between the first two
slepton mass eigenstates.  For a significant portion of the parameter space $\lsptwo$
decays to a muon and a slepton at $50\%$ branching ratio, or one sees the $\tau$ lepton
channel but no electrons, with $50\%$ branching ratio for invisible modes.  Thus the
flavour democratic simplified model scenarios are mostly absent in our model.  As a
result, apart from the invisible modes which have a $50\%$ branching ratio, $\lsptwo$ can
dominantly decay either into $\mu  \tilde{\mu}$ or $\tau  \tilde{\tau}$ with $50\%$
branching ratios. For the first case the usual LHC limit will then put more stringent
bounds.

In the left panel of Fig.~\ref{fig:decmode}, the branching ratios for different leptonic
decay modes are plotted against the $\lsptwo$ mass.  The muon channel is always at $50\%$
branching ratio but the $\tau$ channel has a branching ratio which is mostly less than
$20\%$ with some points having branching ratio in the range between $20-100\%$ while no
electrons are seen.

In Fig.~\ref{fig:chi2}, we present the branching ratios of $\lsptwo$ which decays into a
muon and a slepton (${\tilde{e}_1} / {\tilde{e}_2}$ ), where the slepton further decays
into a muon and an LSP with $100\%$ branching ratio. The points are color coded according
to the mass differences $m_{\tilde{e}}-m_{\lspone}$ (left panel) and
$m_{\lsptwo}-m_{\tilde{e}}$ (right panel).  These BRs($\lsptwo \ra \mu  {\tilde{e}_{1/2}}
\ra \mu \mu \lspone $) vary within $40 - 50 \%$  with the rest being invisible, where the
slepton could be degenerate with either  $\lspone$ or $\lsptwo$.


\begin{figure}[t]
\centering
\subfloat{%
\hspace{-1.2cm}
\includegraphics[width=0.58\textwidth]{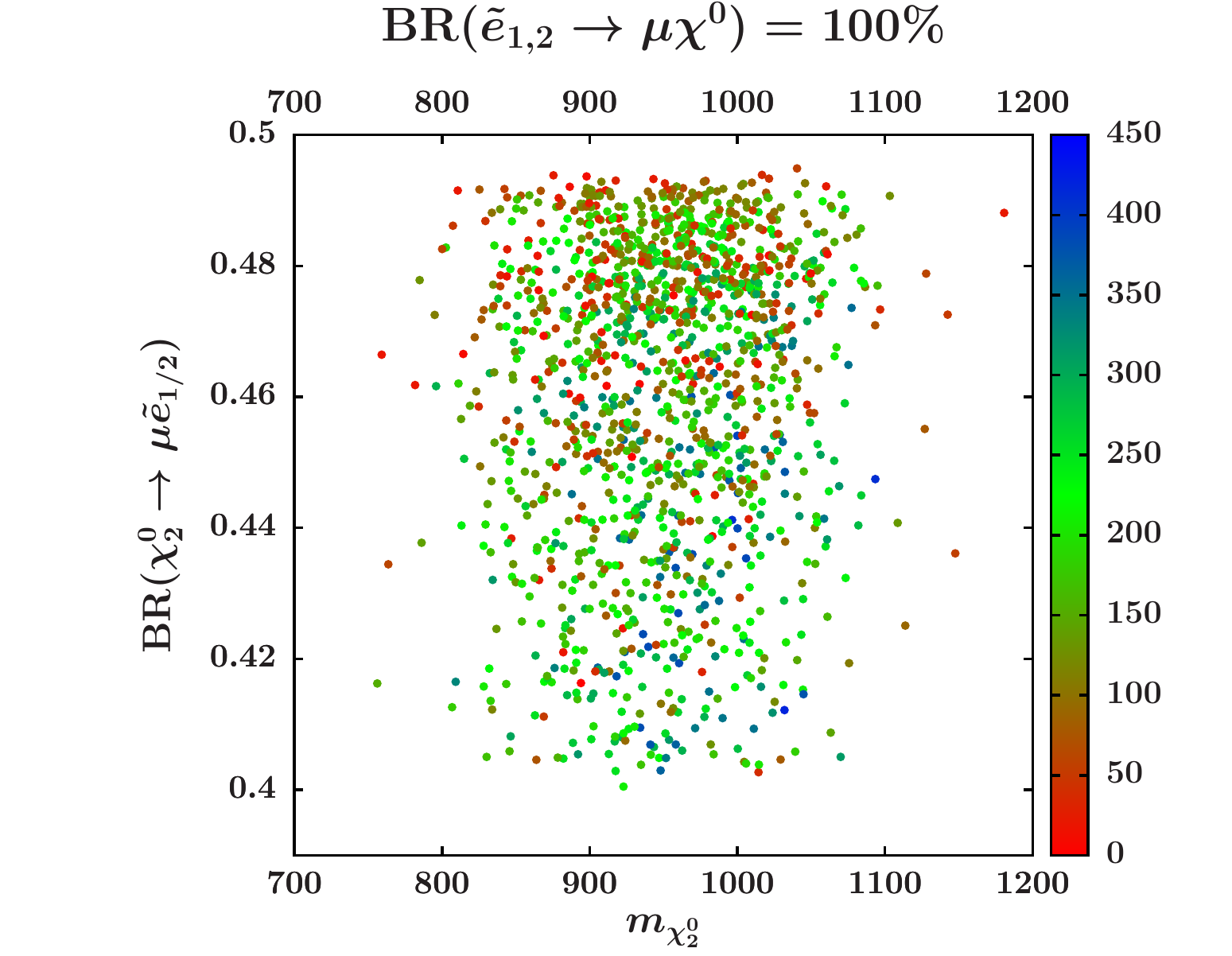}
}%
\subfloat{%
\hspace{-0.08\textwidth}
\includegraphics[width=0.58\textwidth]{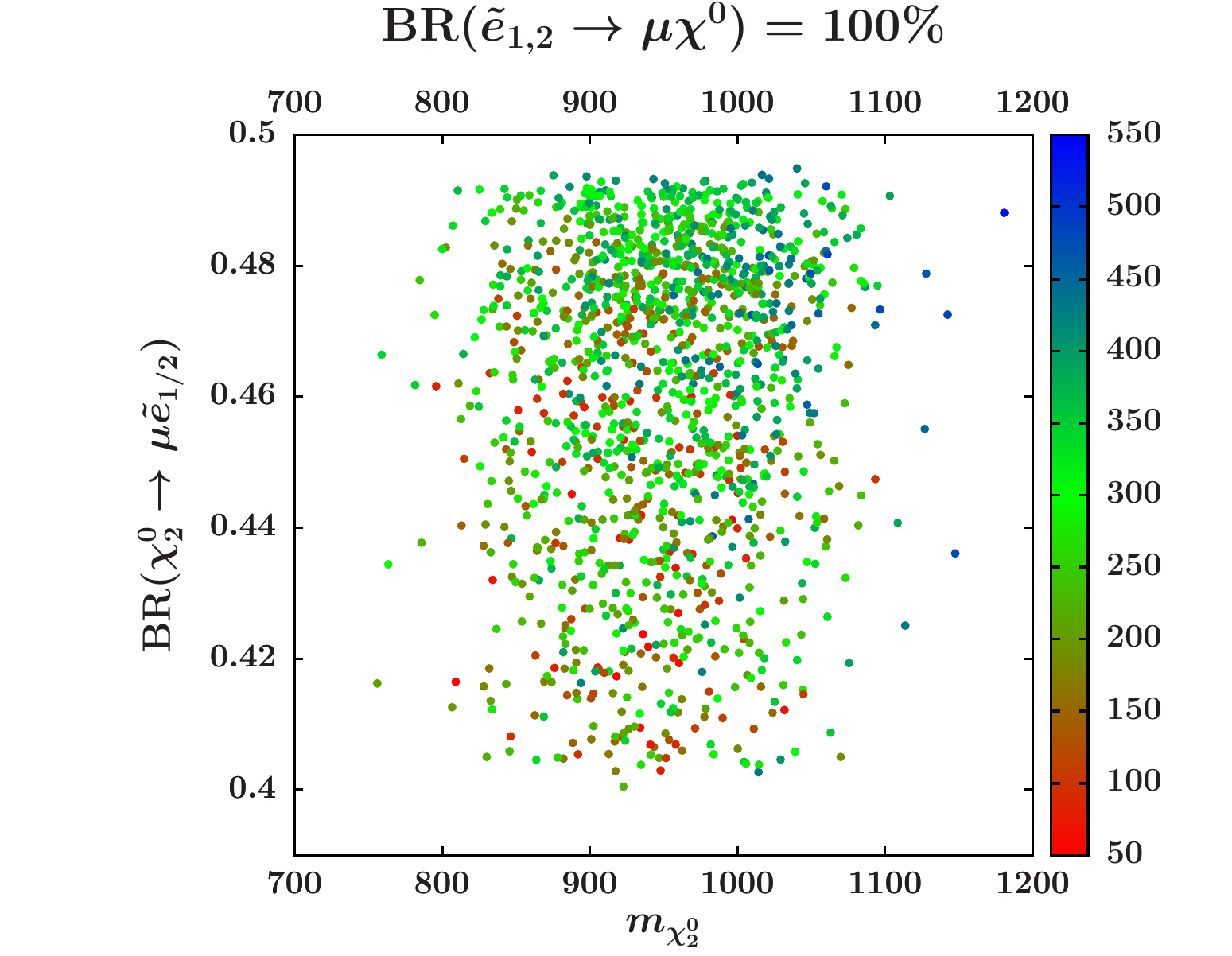}
}%
\caption{Branching ratio of $\lsptwo\to\mu\tilde{e}$ plotted against $\lsptwo$ mass for
parameter space satisfying the constraints described in the text as
well as giving a \gmtwo\  contribution that is within the current $2\sigma$ limit.
The left panel shows the points color coded with slepton-LSP mass difference while
the right panel shows them color coded according to the $\lsptwo$-slepton mass difference. }
\label{fig:chi2}
\end{figure}

\subsection{Decay modes for $\chonepm$:}
\begin{figure}[ht]
\subfloat{%
\hspace{-1.2cm}
\includegraphics[width=0.58\textwidth]{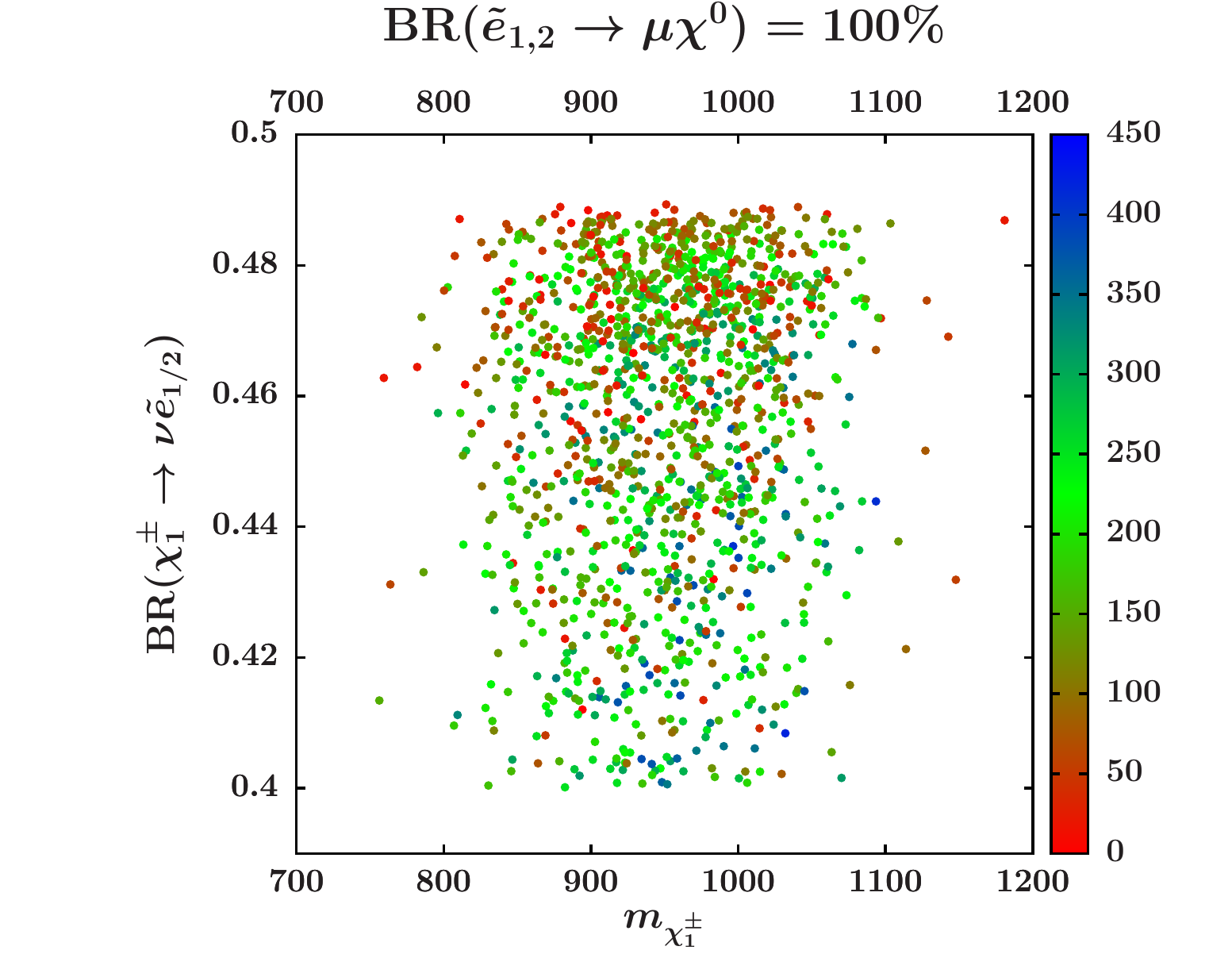}
}%
\subfloat{%
\hspace{-0.08\textwidth}
\includegraphics[width=0.58\textwidth]{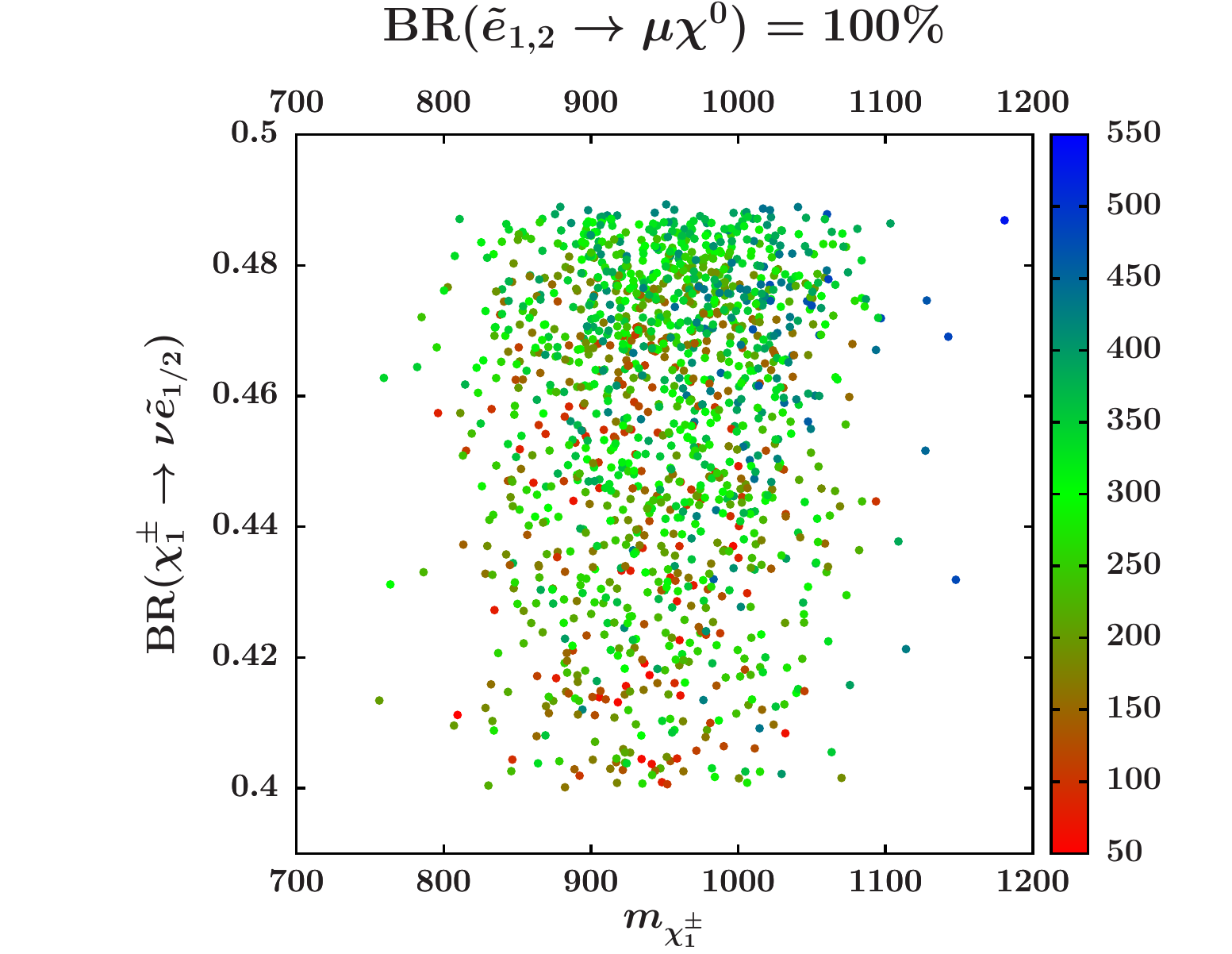}
}%
\caption{Same as Fig.~\ref{fig:chi2} but for $\chonepm\to\nu\tilde{e}$ with the points in
the left panel color coded according to the mass difference $m_{\tilde{e}}-\mlspone$ while
those in the right panel according to $\mchonepm-m_{\tilde{e}}$.}
\label{fig:cha1}
\end{figure}

The charginos decay into $l \tilde{\nu}$ and $\nu \tilde l$ with equal branching ratios
for three generation in ``flavor-democratic'' simplified model. As we discussed in the
previous subsection, due to the different smuon mixing as compared to MSSM, the charginos
largely decay into $\nu \tilde \mu $ and $\mu \tilde{\nu}$  with a branching ratio of 50\%
each or the corresponding $\tau$ lepton channel.
 
In the right panel of Fig.~\ref{fig:decmode}, we show the branching ratios into different
leptonic channels for $\chonepm$ as a function of $\mchonepm$.  We can see that the
chargino dominantly decays to muons with a very small fraction going to $\tau$ and none to
electrons.  Thus a three muon signal is the most likely and also the most constraining
signature to look for in the trilepton searches.

In Fig.~\ref{fig:cha1}, we present the branching ratios of $\chonepm$ where it decays into
a neutrino and a slepton (${\tilde{e}_1} / {\tilde{e}_2}$ ), where the slepton further
decays into a muon and an LSP. The points are color coded according to the mass
differences $m_{\tilde{e}}-\mlspone$(left panel) and $\mchonepm-m_{\tilde{e}}$(right
panel).  These BRs($\chonepm \ra \nu {\tilde{e}_{1/2}} \ra \nu \mu \lspone $) varies
within 40 - 50 \% with the rest being $\chonepm \ra \mu {\tilde{\nu}}$ , where the slepton
could be degenerate with either  $\lspone$ or $\lsptwo$.


\begin{table}[!htbp]
	\begin{center}
		\begin{tabular}{|c|c|c|c|c|}
			\hline
			\hline
			\rule{0pt}{2.5ex}\small{} & Parameter & BP1 & BP2 & BP3 \\
			\hline
			\hline
			\rule{0pt}{2.5ex} & $\mzero$ & $1023$ & $1162$ & $970$ \\
			& $\mhalf$ & $1398$ & $1544$ & $1358$ \\
			& $\azero$ & $36$ & $1317$ & $606$ \\
			& $M_V$   & $329$ & $324$ & $746$ \\
			\small{\textbf{GUT inputs}} & $B_0$ & $692$ & $-410$ &
			$278$ \\
			& $\lambda_5$ & $-0.16$ & $-0.14$ & $-0.16$ \\
			& $\widetilde{m}^2(\times 10^6)$ & $1.6$ & $1.9$ & $1.6$ \\
			& $\widetilde{M}$ & $4.3$ & $4$ & $-10.9$ \\
			\hline
			\rule{0pt}{2.5ex} & $\mathrm{tan}\beta$ & $44.7$ & $48.5$ &
			$48.8$ \\
			& $\lam_{D,2}$ & $-0.39$ & $-0.34$ & $-0.38$ \\
			& $\lam_{L,2}$ & $-0.2$ & $-0.17$ & $-0.19$ \\
			\hline
			\rule{0pt}{2.5ex} & $m_{h}$ & $124$ & $123$ & $123$ \\
			& $m_{\chi_1^0}$ & $474$ & $526$ & $463$ \\
			& $m_{\charone}$ & $898$ & $993$ & $875$ \\
			\small{\textbf{Pole masses}} & $m_{\tilde{e}_1}$ & $484$ &  $576$
			& $669$ \\
			& $m_{\tilde{e}_2}$ & $858$ &  $866$ & $752$ \\
			& $m_{\tilde{\nu}_1}$ & $475$ &  $569$ & $663$ \\
			& $m_{\tilde{t}_R}$ & $2021$ &  $2297$ & $1986$ \\
			\hline
			\hline
			& $\charone\to\mu{\tilde{\nu}}$ & $0.5$ & $0.5$ & $0.5$ \\
			& $\charone\to\nu{\tilde{e}_1}$ & $0.49$ & $0.48$ &
			$0.47$ \\
			\small{\textbf{Branching Ratios} } & $\chi_2^0\to\mu{\tilde{e}_1}$ &
			$0.49$ & $0.49$ & $0.48$ \\
			& $\chi_2^0\to\nu{\tilde{\nu}}$ & $0.5$ & $0.5$ &
			$0.49$ \\
			& ${\tilde{e}_1}\to\mu\chi_1^0$ & $1.0$ & $1.0$ & $1.0$ \\
			\hline
			& $\delta\gmtwo(\times 10^{-9})$ & $2.54$ & $2.23$ & $2.09$\\
			& $\Delta m=m_{\tilde{e}_1}-m_{\chi_1^0}$ & $10$ & $50$ &
			$\left.\left(m_{\chi_2^0} - m_{\chi_1^0}\right)\middle/2\right.$ \\
			\hline
			\hline
		\end{tabular}
		\caption{Benchmark points chosen such that they satisfy the constraints as
		described in the text as well as giving a contribution to \gmtwo\  that is
		consistent with the current $2\sigma$ limit.  All masses are in GeV.}
		\label{tab:benchm} 
	\end{center}
\end{table}

\subsection{Benchmark points and models}

From the results of the previous section on the decay modes of $\chonepm$ and $\lsptwo$,
we can see that the collider constraint that is best suited to probe the
chargino-neutralino pair production are the trilepton searches.  In
Table~\ref{tab:benchm}, we show three benchmark points chosen from Figs.~2-4.   The decay
properties of these points are strikingly different from the usual simplified models
considered by LHC collaboration to interpret the trilepton limits.  Also the mass
hierarchies between sleptons and the electroweakinos are different in our model.
Motivated by these benchmark points we choose the following three scenarios, or benchmark
models.

\bi 
\item 

\textbf{Benchmark Model 1 (BM1)}: This model is motivated by benchmark point 1 (BP1) where
the electroweakinos dominantly decay into muons. Here sleptons are NLSP and nearly
degenerate with the LSP and we assume  $m_{\tilde{\nu}_1}$ = $m_{\tilde{e}_1}$ =
$m_{\chi_1^0}$ + 10 GeV.  For the branching ratios of the electroweakinos we assume
BR$(\charone \to$  $\mu {\tilde{\nu}_1},\,\nu {\tilde{e}_1})=0.50$, and BR$(\chi_2^0\to
\mu {\tilde{e}_1},\,\nu {\tilde{\nu}})=0.50$; where BR$({\tilde{e}_1}$ $\to \mu
{\chi_1^0})=1.0$. These assumptions apply to each point in the parameter space.

\item \textbf{Benchmark Model 2 (BM2)}: BM2 is motivated by BP2 and the decay patterns of
	BM2 are same as BM1 but with the choice $m_{\tilde{\nu}_1}$ = $m_{\tilde{e}_1}$ =
	$m_{\chi_1^0}$ + 50 GeV. This choice of mass dependence can significantly change
	the limits on chargino masses.

\item \textbf{Benchmark Model 3 (BM3)}: BM3 is motivated by BP3 and the decay patterns of
	chargino and neutralino are similar to previous benchmark models. We choose the
	slepton mass as: $m_{\tilde{\nu}_1}$ = $m_{\tilde{e}_1}$ = ($m_{\chi_2^0}$ +
	$m_{\chi_1^0}$)/2.  This choice of mass basically is similar to the simplified
	models considered by ATLAS, but BM3 is different in terms of branching ratios.

\ei

\section{Collider analysis for trilepton searches\label{trilepton}}

Both CMS and ATLAS Collaborations have searched for the EWinos with different final states
($2l, 3l,$ with/without $taus, lbb, l\gamma\gamma$ etc.) from direct pair production of
$\chonepm \lsptwo$ or $\chonepm
\chonepm$\cite{Aad:2014nua,Aad:2014vma,Aad:2014yka,Aad:2015jqa,Aad:2015eda,
Khachatryan:2014mma,Khachatryan:2014qwa,atlas_ew_3l,CMS-PAS-SUS-17-004}.  The results are
mainly interpreted for \textit {Slepton mediated simplified model}, \textit {WZ mediated
simplified model} and \textit {Wh mediated simplified model}. In the first case, the
sleptons are assumed to be lighter than $\chonepm$ and $\lsptwo$ and this channel gives
the most stringent bounds as the EWinos decay via slepton to lepton enriched final
states\cite{atlas_ew_3l}.  For rest of the two cases, sleptons are assumed to be much
heavier than $\chonepm$ or $\lsptwo$ and the electroweakinos decay via real or virtual
$W$, $Z$ and Higgs boson. In our own model, the LHC limits on gluinos from 13 TeV data
put stringent bounds on $\mchonepm \gtrsim 700$ GeV (due to High scale input) and only the
trilepton analysis targeting  $\chonepm \lsptwo$ production is sensitive to $\mchonepm >
700$ GeV region.  Hence in this analysis we only focus on the trilepton channels
(dedicated signal regions for  \textit {Slepton mediated simplified model}).  First we
will briefly discuss about the 13 TeV trilepton search analysis considered by
ATLAS\cite{atlas_ew_3l} and present our results alongside ATLAS for validation and direct
comparison.

\begin{table}[t]
\begin{center}
\begin{tabular}{||c||c|c||c||c|c||}
\hline
\hline
	 			&SR$3l$-a   &  SR$3l$-b  &SR$3l$-c   &  SR$3l$-d  &SR$3l$-e   		\\
\hline 
$N_{lepton}$		& \multicolumn{5}{c||}{3} 				\\
\hline
$\met >$ (GeV)	& \multicolumn{5}{c||}{130} 				\\
\hline
$m_{T}^{min} > (GeV)$	& \multicolumn{5}{c||}{110} 				\\
\hline
$m_{SFOS}$	(GeV)	& \multicolumn{2}{c||}{$< 81.2$}	&  \multicolumn{3}{c||}{$> 101.2$}\\
\hline
$p_{T}^{l_3}$ (GeV)	 &  	20 - 30		 &	$>$ 30		  &  20-50 		& 50-80		&	$>$ 80	\\
\hline

\hline
Observed events 	 &  		4	 &3			  &9  			&0		&	0	\\
\hline 
Total SM	 &  	2.23 $\pm$ 0.79 &	2.79$\pm$0.43  & 5.41 $\pm$ 0.93	&1.42 $\pm$ 0.38&1.14 $\pm$ 0.23	\\
\hline
\hline
       \end{tabular}\
       \end{center}
           \caption{Selection requirements for slepton mediated ($3l$) channel considered by ATLAS 
       for 13 TeV 36.1 $\ifb$ data \cite{atlas_ew_3l}.}
\label{tab:cuts}
\end{table}

\subsection{Validation for trilepton analysis}

\begin{figure}[bt]
	\centering
		\includegraphics[width=0.7\textwidth]{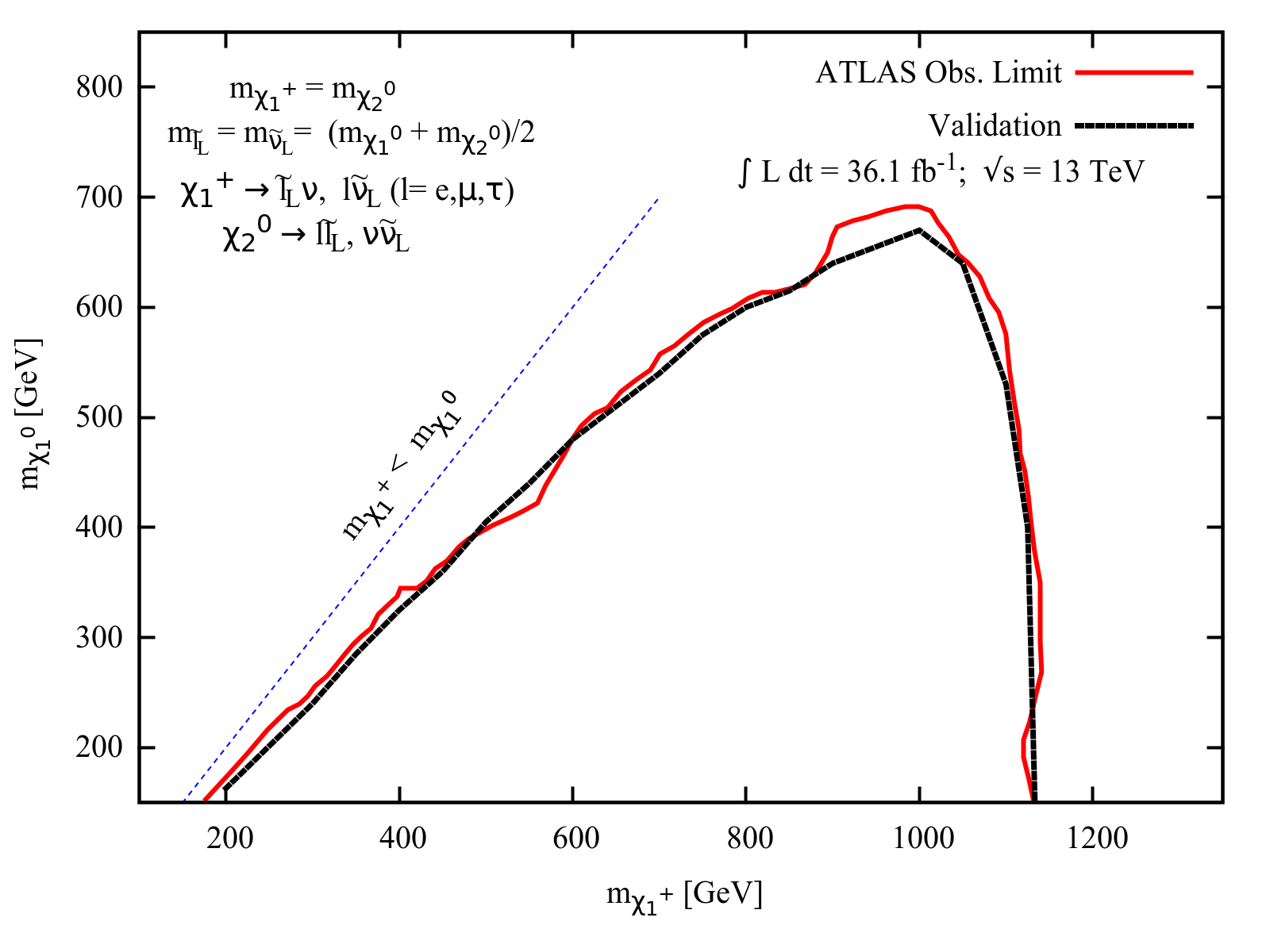}
		\caption{Validation of the ATLAS trilepton analysis for Run-II 36.1 $\ifb$ 
		data\cite{atlas_ew_3l}.
	The exclusion limit in the $\mlspone$ - $\mchonepm$ plane obtained by the ATLAS
Collaboration (red line) in their trilepton analysis is reproduced using similar mass 
relations and branching ratios of the relevant gauginos and sleptons (black line).} 
	\label{fig:3l_slepton}
\end{figure}
 
 
In \textit{slepton ($\wt\ell_L$)-mediated} models, it is assumed that the left handed
sleptons and sneutrinos lie exactly midway between $\lspone$ and $\lsptwo$,
$m_{\wt\ell_L}$ = ($\mchonemp + \mlsptwo$)/2, and the EWinos decay either to left handed
sleptons or sneutrinos universally.  Events are considered with exactly three tagged
leptons (electron or muon) \cite{atlas_ew_3l}.   Event reconstruction details like
electron, muon, tau and jet identification, isolation, overlap removal etc. are followed
according to the ATLAS analysis as mentioned in Sec. 5 and Sec. 6 of \cite{atlas_ew_3l}.
In this trilepton analysis,  a veto on $b$-jet is applied to all signal channels.  For
$b$-jets, we use the $p_T$ dependent $b$-tagging efficiencies obtained by ATLAS
collaboration in Ref.\cite{ATLAS-CONF-2012-043}. 

\begin{figure}[t]
	\centering
		\includegraphics[width=0.7\textwidth]{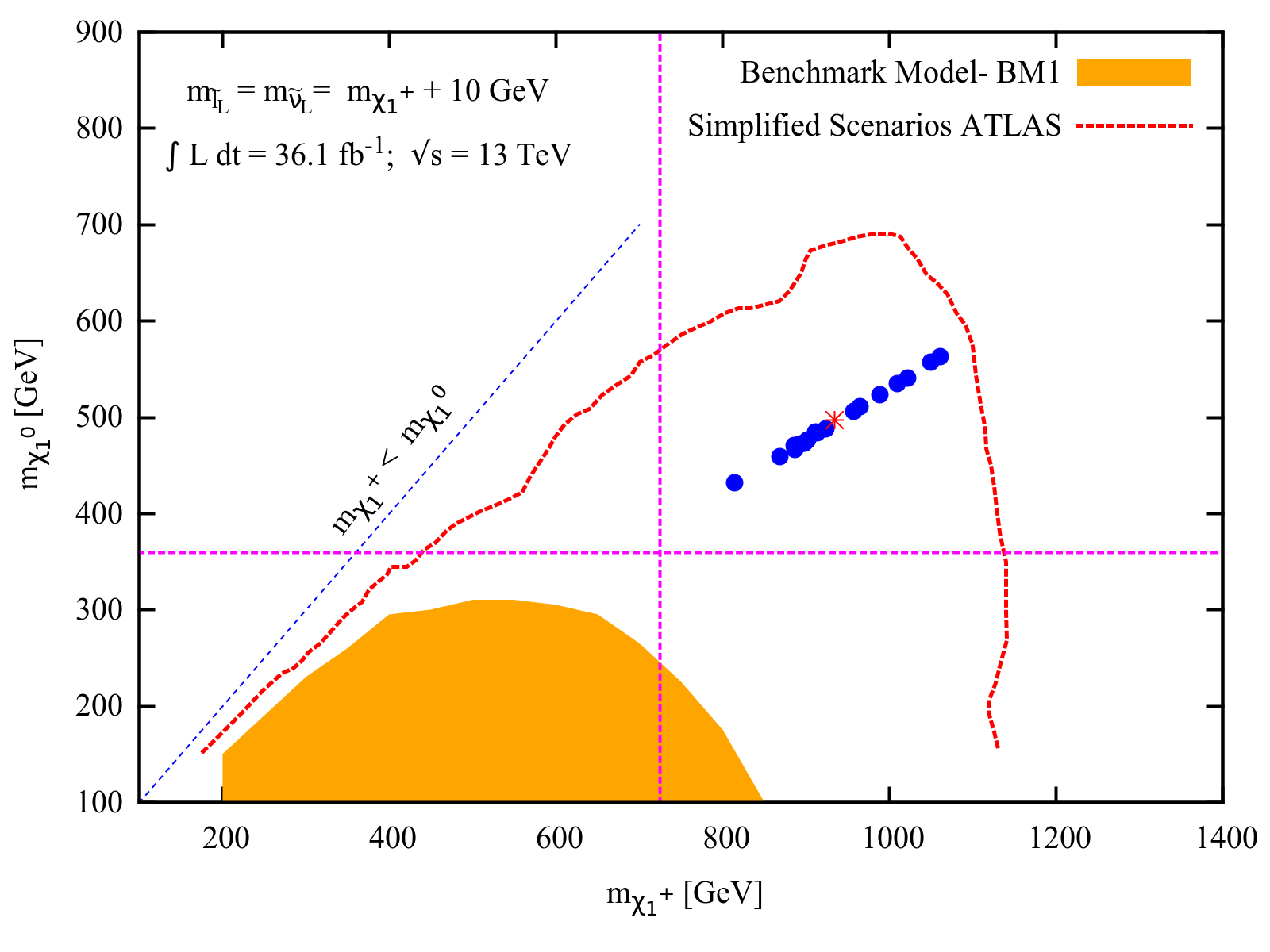}
	\caption{Limits on $\mchonepm-\mlspone$ plane for Benchmark Model 1 (see text).  The excluded region 
	for BM1  is shown in orange.  The red dotted line represents the limit for slepton mediated 
	simplified scenarios\cite{atlas_ew_3l}. The magenta lines
        indicate the indirect lower limit on $\mchonepm$ (vertical line) and
$\mlspone$ (horizontal line) from gluino seaches in 13 TeV
data\cite{ATLAS-CONF-2017-022}.  The blue points (circle) will be 
allowed by the New Muon $g-2$ result in a future measurement at Fermilab\cite{Grange:2015fou,Chapelain:2017syu} while the red
points (star) will be ruled out, assuming that the central value of the measurement remains
the same. }
	\label{fig:bm1}
\end{figure}

Depending upon the requirement of $m_{SFOS}$ (invariant mass of same-flavour opposite-sign
(SFOS) lepton) and  $p_{T}^{l_3}$ ($p_{T}$ of third leading lepton),  ATLAS has optimised
five signal regions (SR), namely, SR$3l$-a to SR$3l$-e for \textit {Slepton mediated
simplified model}. The basic selection requirements for these channels, number of observed
events and total SM background are listed in Table~\ref{tab:cuts}. In the absence of any
BSM signal in all these channels, limits are set  on the number of SUSY signal events at
95\% confidence level (CL).  For these five signal regions (SR$3l$-a to SR$3l$-e)
$N_{\rm{BSM}}$ at  95\% CL are 7.2, 5.5, 10.6, 	3.0 and 3.0, respectively.  The ATLAS
Collaboration has translated these obtained upper limits on $N_{\rm{BSM}}$ into exclusion
limits in the $\mlspone$ - $\mchonepm$ plane.  In a similar way, we have also reproduced
the exclusion contours obtained by ATLAS assuming similar mass relations and branching
ratios of the relevant gauginos and sleptons.  In order to validate our results we
reproduce the exclusion contours using PYTHIA (v6.428)\cite{Sjostrand:2006za}
\footnote{These same set-up of codes were also used in Ref.\cite{Choudhury:2016lku}}.  We
use the next-to-leading order (NLO) + next-to-leading logarithmic (NLL)
chargino-neutralino pair production cross-sections given in Ref.\cite{LHCSXSECWG}, which
have been calculated for 13 TeV using the resummino code\cite{Fuks:2012qx,Fuks:2013vua}.
For \textit {slepton mediated models}, SR$3l$-e is the most sensitive channel for the
parameter space with large mass splitting between $\chonepm$ and $\lspone$ ($\delta m = $
$\mchonepm - \mlspone$).  For smallest $\delta m$, low-valued $m_{SFOS}$ SR$3l$-a  is more
effective to probe the relevant parameter space.


In Fig.~\ref{fig:3l_slepton},  we present the validated results for  \textit {slepton
mediated simplified models}.  The red line corresponds to 95 \% CL exclusion limits
obtained by ATLAS and the black line corresponds to our validated results adopting the
ATLAS analysis. From Fig.~\ref{fig:3l_slepton},  it is evident that our validated results
are in good agreement with that of ATLAS.

\subsection{New limits for benchmark models}

First we present our results for \textit{Benchmark Model~1} (\textbf{BM1}), where the
scenarios represent slepton co-annihilation regions and due to the extreme mass degeneracy
($\delta m = $ $\mchonepm - \mlspone$ = 10 GeV) the leptons coming from slepton decay are
very soft (below the trigger cuts).  The soft leptons cause the reduction on limits on
chargino masses. The orange regions in Fig.~\ref{fig:bm1} are excluded from 13 TeV data
where the red dotted line represents the usual limits from simplified scenarios with the
sleptons being midway between LSP and charginos. The vertical and horizontal magenta lines
present the indirect limit on $\mchonepm$ and $\mlspone$ from the gluino limits coming
from 13 TeV data\cite{ATLAS-CONF-2017-022}.  The blue points (circle) are allowed by the
New Muon $g-2$ result in a future measument at
Fermilab\cite{Grange:2015fou,Chapelain:2017syu} while the red points (star) are ruled out,
assuming that the central value of the measurement remains the same.  It is clear from
Fig.~\ref{fig:bm1} that for co-annihilation scenarios trilepton limits are even weaker
than the indirect bounds from direct gluino searches due to the mass correlations in GUT
models.  It may be noted that with non-universal gaugino mass models the indirect limits
from gluino searches are not valid and the models have a wide range of parameter space
which are still allowed (except for the magenta regions).

\begin{figure}[t]
	\centering
		\includegraphics[width=0.7\textwidth]{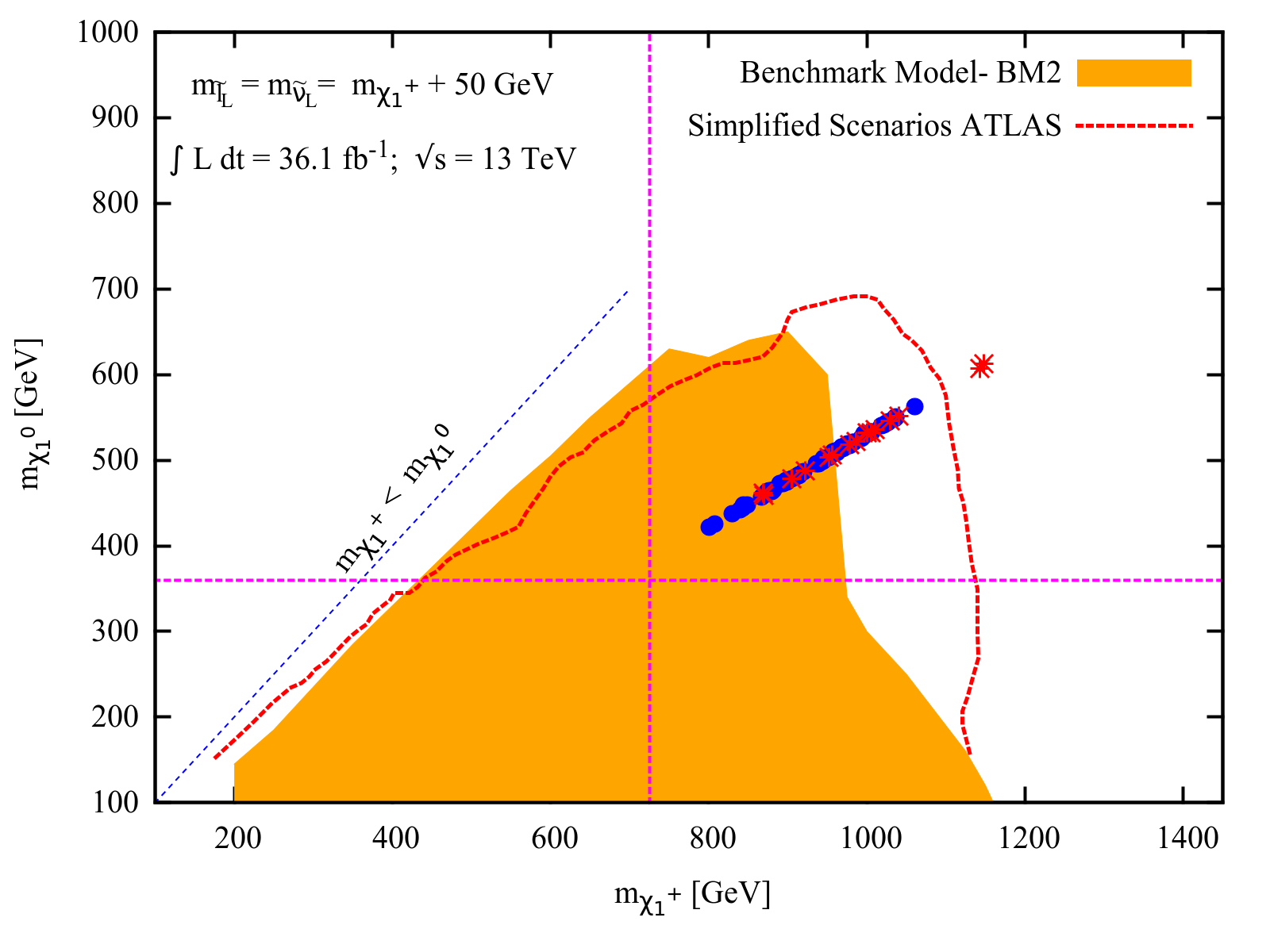}
		\caption{Same as Fig.~\ref{fig:bm1} but for Benchmark Model 2 (see text).}
	\label{fig:bm2}
\end{figure}

The situation changes drastically if the mass difference is somewhat larger. We present
the implication of trilepton data in Fig.~\ref{fig:bm2} for \textbf{BM2} where the mass
splitting between $\chonepm$ and $\lspone$, $\delta m = $ $\mchonepm - \mlspone$, is 50
GeV. The orange regions in Fig.~\ref{fig:bm2} are excluded for \textbf{BM2}.  In some
region of the parameter space the limits are stronger than usual simplified models
(denoted by red line). This is simply due to the enhancement of branching ratios in our
model. It may be noted that for the  simplified models considered by ATLAS, the
electroweakinos decay to leptonic final state universally (but for \textbf{BM2},
electroweakinos decay mainly to $\mu$ final states). In Fig.~\ref{fig:bm2}, roughly half
the points outside the orange shaded region are ruled out by Future Muon $g-2$
experiment\cite{Grange:2015fou,Chapelain:2017syu} as indicated by the red points.

\begin{figure}[t]
	\centering
		\includegraphics[width=0.7\textwidth]{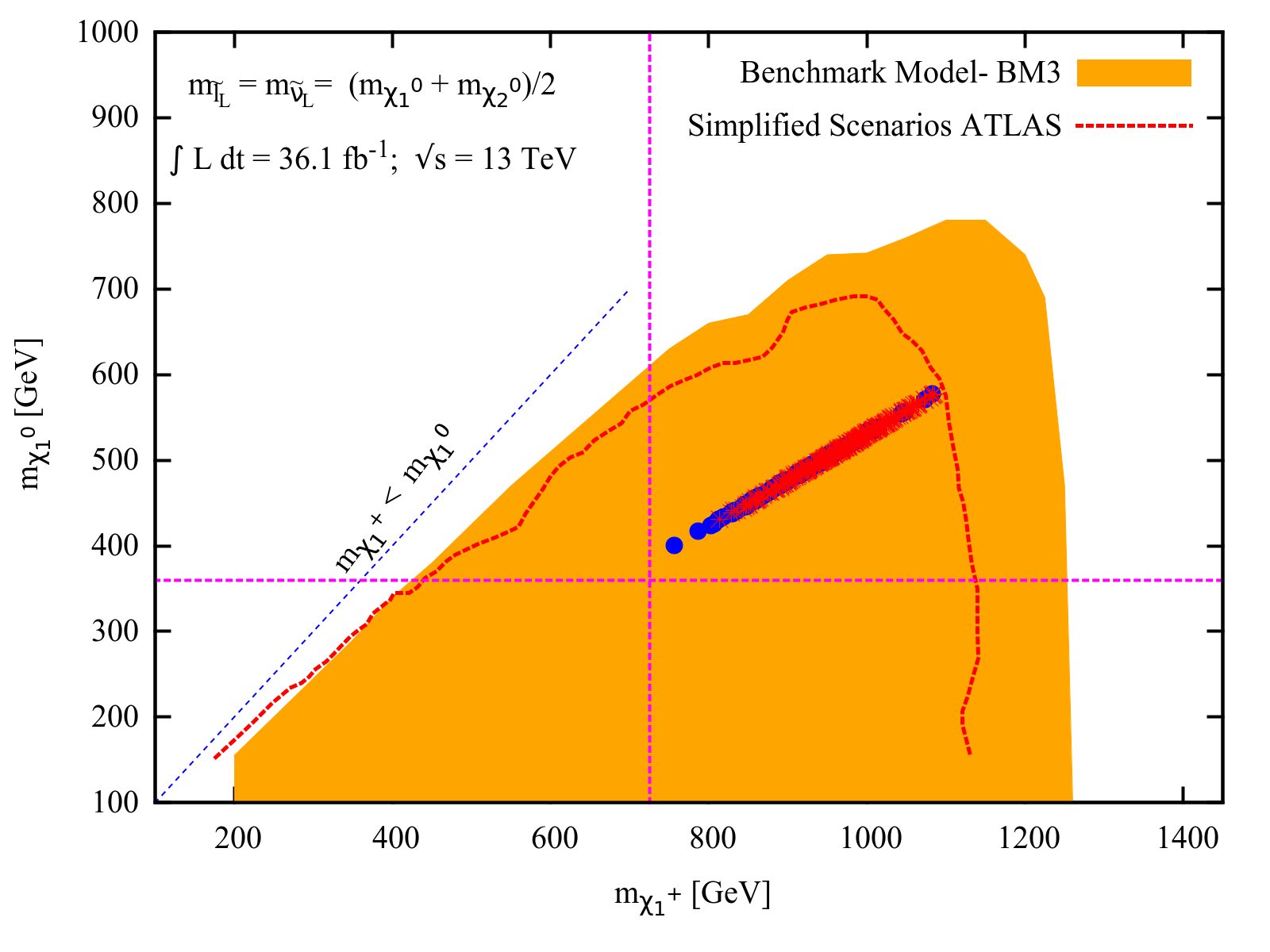}

		\caption{Same as Fig.~\ref{fig:bm1} but for Benchmark Model 3 (see text).}
	\label{fig:bm3}
\end{figure}

In Fig.~\ref{fig:bm3} we analyse the model \textbf{BM3} where the sleptons are exactly
midway of $\chonepm$ and $\lsptwo$ (same choice as simplified models). Similar to
\textbf{BM1}  and \textbf{BM2},  EWinos decay also mainly to $\mu$ final states in
\textbf{BM3}.  For this model, the limits are even stronger than \textbf{BM2}.  For light
$\lspone$, the limit on chargino mass extends upto 1250 GeV.  Hence the current LHC data
exclude all the \gmtwo\ allowed points which have the same characteristic like
\textbf{BM3}.

\section{Conclusions\label{conclusions}}

In this work we have studied the VL extension of MSSM -- by the addition of a pair
$\mathbf{5}+\mathbf{\bar{5}}$ of $SU(5)$ which leads to an additional quark, lepton and a
pair of neutrinos with corresponding squarks, sleptons and sneutrinos -- in the context of
\gmtwo, various flavor physics constraints, DM constraints and LHC limits on squarks and
gluinos.  We identify that the allowed parameter space in Fig.~1 leads to chargino mass in
the range of $700-1200$ GeV.  The mixing of the second and third generation leptons with
the extended spectrum of VL particles leads not only to an enhanced contribution to
\gmtwo\ but also gives a very different kind of signature for electroweakino decay modes.

To probe the allowed parameter space at the LHC, the most sensitive search will be the
trilepton signal coming from chargino-neutralino pair production.  For this reason we do a
detailed study of relevant decay properties and the mass hierarchies in Section 4.1 and
4.2.  In particular we observe that the VL extension of MSSM along with the specific
choice of GUT scale parameters made here leads to a 3 muon or 3 tau final state instead of
lepton universality assumed in the LHC trilepton analysis.  We therefore recast the ATLAS
trilepton searches in chargino-neutralino pair production using the recent Run-II data.
We identify three benchmark points from the scanned dataset. To interpret the trilepton
search we construct three simplified benchmark models based on these benchmark points.  We
observe that the slepton coannihilation scenario, i.e. BM1, is not at all sensitive to the
current LHC data due to the soft nature of the lepton signal (see Fig.~6).  However, the
points with a relatively larger mass difference, as for example BM2, can exclude chargino
mass up to 1 TeV (see Fig.~7).  The strongest constraint comes from BM3 where the slepton
mass lies midway between the chargino and second lightest neutralino.  For such a choice
any parameter range allowed by \gmtwo\ data is already excluded (see Fig.~8).  There still
exists more than half of the parameter space that is not covered by the three benchmark
models considered here, which were chosen such that they are most sensitive to the
trilepton searches.  For this parameter space, the allowed $2\,\sigma$ range of
\deltagmtwomu\ from the New Muon $g-2$ experiment at
Fermilab\cite{Grange:2015fou,Chapelain:2017syu} can potentially rule out more than two
thirds of the region, assuming that the central value of the \gmtwo\ measurement remains
the same.  Much of this parameter space consists of tau lepton final states in
chargino-neutralino pair production, which is not sensitive to the current Run-II data
with $\lum = 36.1 \ifb$\cite{ATLAS-CONF-2017-035}.  Future searches with higher luminosity
for $2\tau$/$3\tau$ signal at the LHC could potentially probe this region of the parameter
space.

\bigskip

\noindent \textbf{Acknowledgments}
\medskip

\noindent 
We would like to thank  Luc Darm\'e and Enrico Maria Sessolo for the useful discussions in
the initial stage of this work. AC and SR would like to thank Luc Darm\'e for helpful
inputs to numerical computation. AC and LR are supported by the
Lancaster-Manchester-Sheffield Consortium for Fundamental Physics under STFC Grant No.\
ST/L000520/1. SR and LR are supported in part by the National Science Council (NCN)
research grant No.~2015-18-A-ST2-00748.  The use of the CIS computer cluster at the
National Centre for Nuclear Research in Warsaw is gratefully acknowledged.

\bibliographystyle{JHEP}

\bibliography{VL}

\end{document}